\documentclass{aa}  

\usepackage{graphicx}
\usepackage{natbib}
\usepackage{xcolor}
\usepackage{multirow}
\usepackage{ulem}
\usepackage{CJKutf8}
\usepackage{comment}
\usepackage{makecell}
\usepackage{txfonts}
\usepackage{amsmath}
\usepackage{placeins} 

\usepackage{nccmath}
\usepackage[colorlinks=true,linkcolor=blue, citecolor=blue]{hyperref}%

\newcommand{\tweco}{$^{12}$CO($J$=1$\rightarrow$0)}
\newcommand{\thico}{$^{13}$CO($J$=1$\rightarrow$0)}
\newcommand{\eigco}{C$^{18}$O($J$=1$\rightarrow$0)}
\newcommand{\hcn}{HCN($J$=1$\rightarrow$0)}
\newcommand{\hnc}{HNC($J$=1$\rightarrow$0)}
\newcommand{\so}{SO($J_K$=3$_2$$\rightarrow$2$_1$)}
\newcommand{\cs}{$^{12}$CS($J$=2$\rightarrow$1)}
\newcommand{\hcop}{HCO$^+$($J$=1$\rightarrow$0)}
\newcommand{\nhp}{N$_2$H$^+$($J$=1$\rightarrow$0)}

\newcommand{\NH}{$N_{\rm{H_2}}$}

\newcommand*\diff{\mathop{}\!\mathrm{d}}
\newcommand{\modifi}[1]{\textcolor{black}{#1}}
\newcommand{\modifj}[1]{\textcolor{black}{#1}}
\newcommand{\modifr}[1]{\textcolor{black}{#1}}
\newcommand{\modifrr}[1]{\textcolor{black}{#1}}
\newcommand{\modifrrr}[1]{\textcolor{black}{#1}}

\newcommand{\FigEmissionFunction}{%
    \begin{figure}
        \centering %
        \includegraphics[width = 1\linewidth]{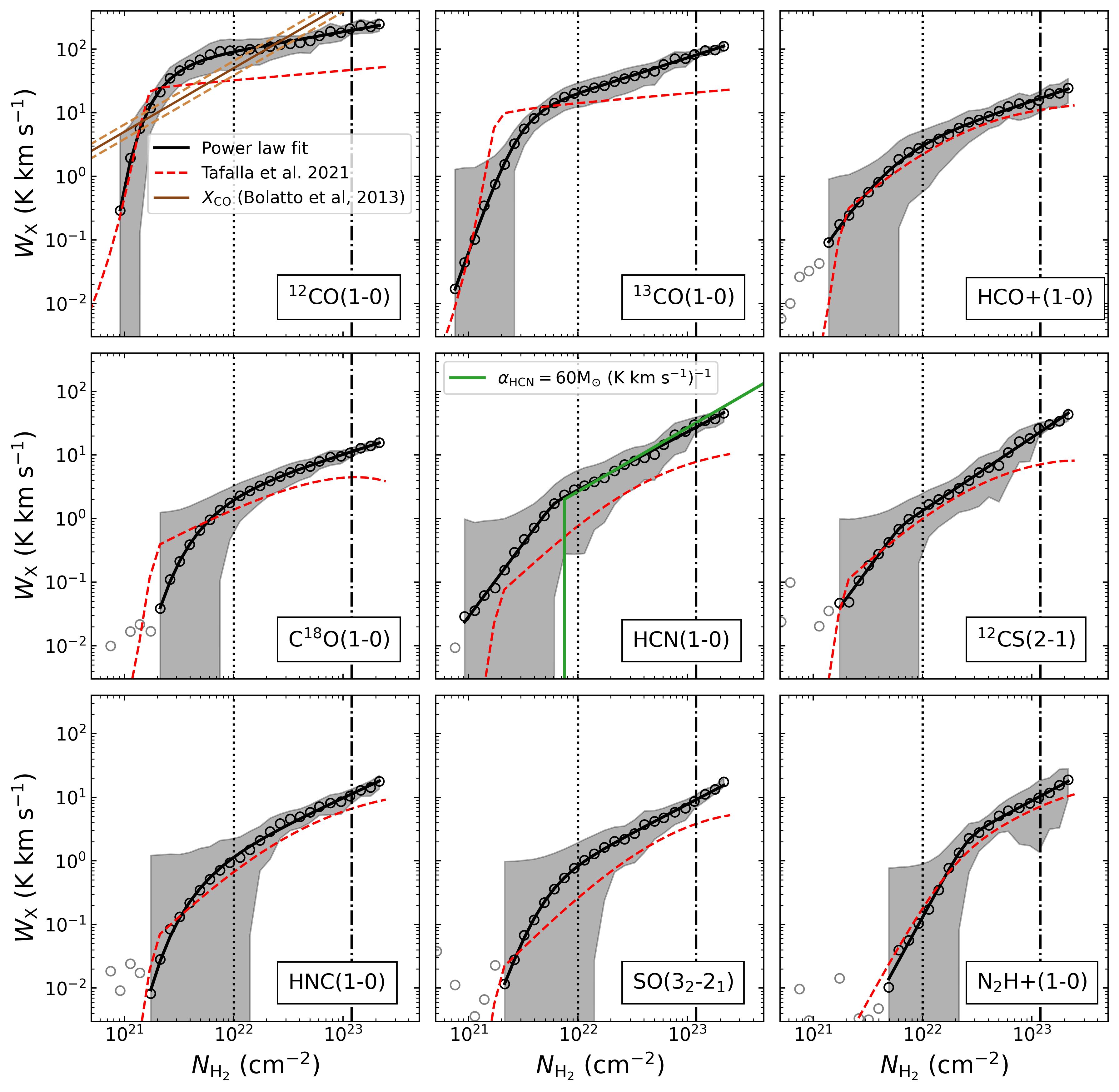}
    \caption{Binned trends of line integrated intensity as a function of column density. The data is binned in 30 equally sized bins of column density. Black circles correspond to the bin average, while the grey shading indicates the standard deviation in each bin. The black solid line is a smoothly varying double PL fit to the trends, specific to each emission line. The red dashed line shows for comparison the empirical fit to \modifr{the Perseus cloud} by \cite{Tafalla2021}, assuming a kinetic temperature of 11\,K. Each panel shows a different emission line: \tweco{}, \thico{} and \hcop{} (left to right, top row);  \eigco{}, \hcn{} and \cs{} (middle); \hnc{} \so{} and \nhp{} (bottom). The standard Milky Way CO-to-H$_2$ conversion factor and its typical uncertainty \citep{Bolatto2013} is indicated in the top left panel. An \hcn{} dense gas conversion factor of 60 M$_\odot$ (K km s$^{-1}$)$^{-1}$ is indicated in the central panel \modifr{as the green curve}. }
    \label{fig:emission-function}
    \end{figure}
}

\newcommand{\PostPDFExampleCorr}{%
    \begin{figure}
        \centering %
        \includegraphics[width = 1\linewidth]{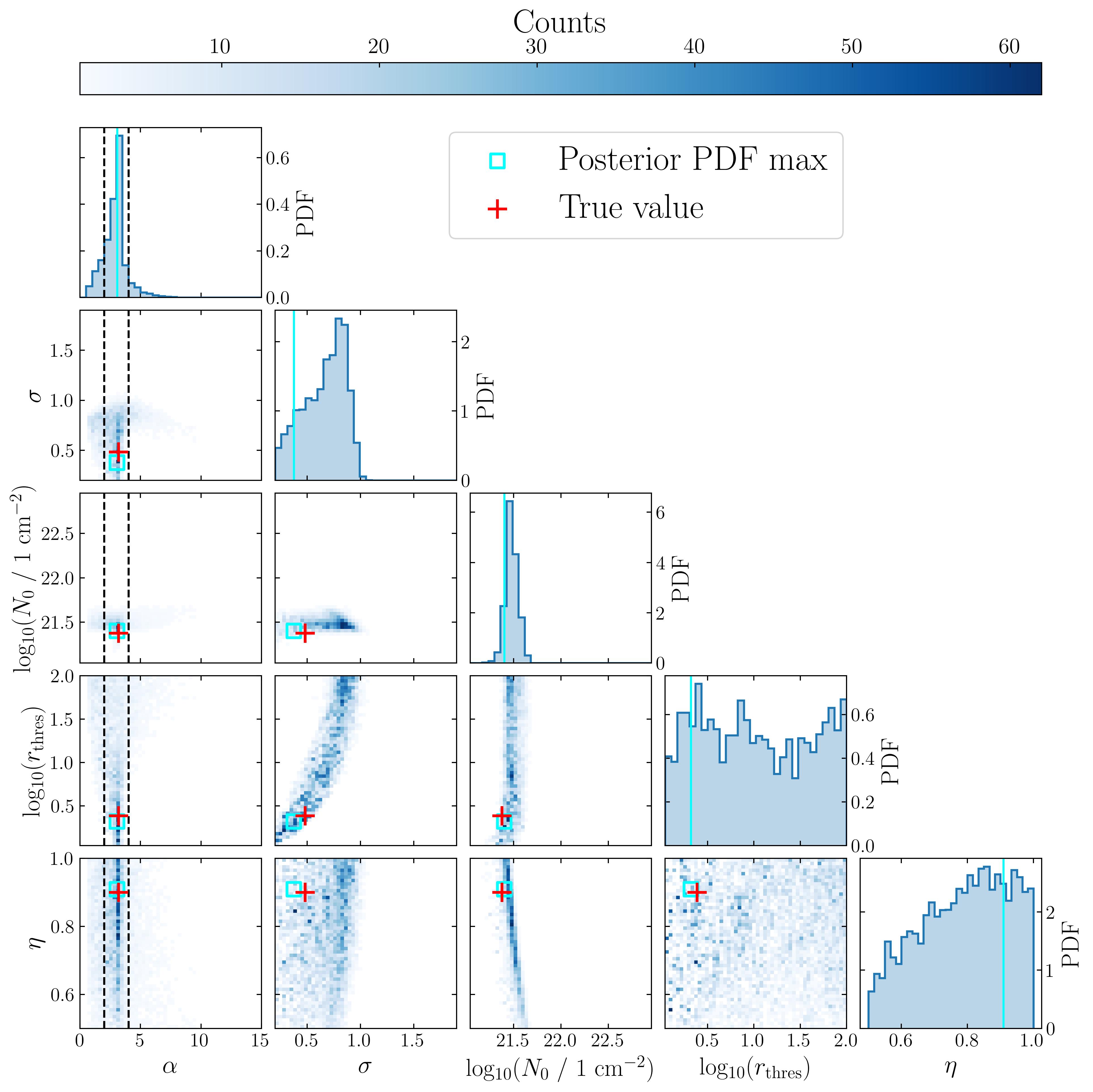}
    \caption{Two-dimensional projections of the posterior PDF in the form of a scatter plot matrix. The matrix's diagonal shows the posterior PDF of each estimated parameter. The MAP estimation is represented as a vertical cyan line on the histograms and as a cyan square in the scatter plot. The true $N$-PDF parameters obtained by fitting the dust derived Orion B $N$-PDF is shown are shown as red crosses. The black dashed line show the range in PL index $\alpha$ of the $N$-PDF expected for gravitational collapse. The estimations closely match the reference values, although clear degeneracies are present in the posterior PDF. }
    
    \label{fig:orionb-post-pdf-corr}
    \end{figure}
}

\newcommand{\PostPDFExample}{%
    \begin{figure}
        \centering %
        \includegraphics[width = 0.8\linewidth]{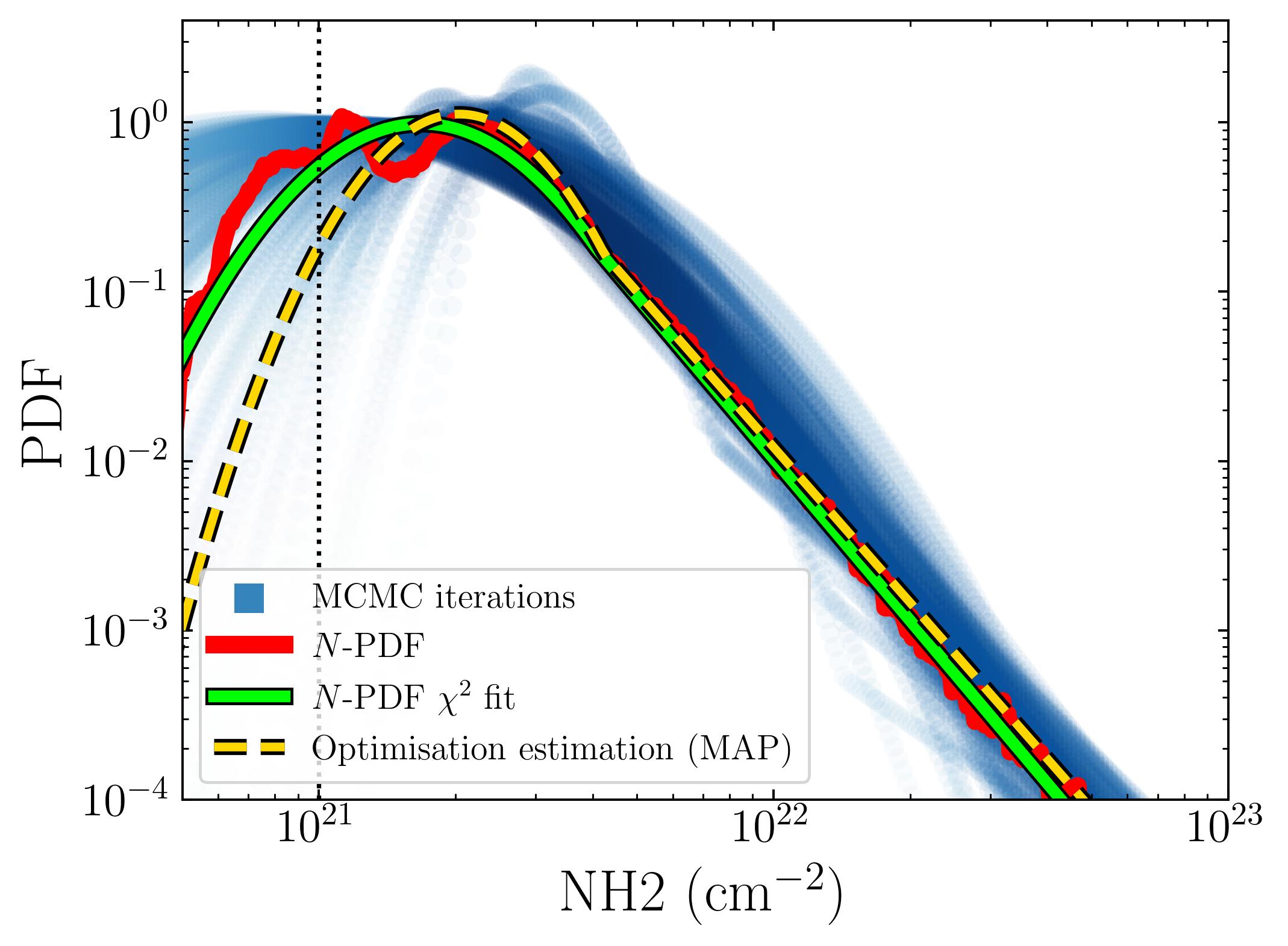}
    \caption{A comparison of the reference and estimated
     $N$-PDFs when inverting the $N$-PDF on the spatially and spectrally
     averaged ORION-B data. The thick red line indicates the $N$-PDF as a histogram constructed directly from the dust-derived Orion B column densities, and the green line represents a $\chi^2$ fit to the red histogram. The estimated $N$-PDFs from the 10\,000 MCMC iterations to sample the Bayesian posterior are shown with blue
     circles. The dashed orange line is the MAP estimation for the $N$-PDF. The vertical dotted black line indicates the limit below which the line intensities
     predicted by the emission function fall below the typical noise level
     of the data, that is 0.1 K\,km\,s$^{-1}$.}
    \label{fig:otrionb-post-pdf}
    \end{figure}
}

\newcommand{\ParamsMap}{%
    \begin{figure*}
        \centering %
        \includegraphics[width = 0.8\linewidth]{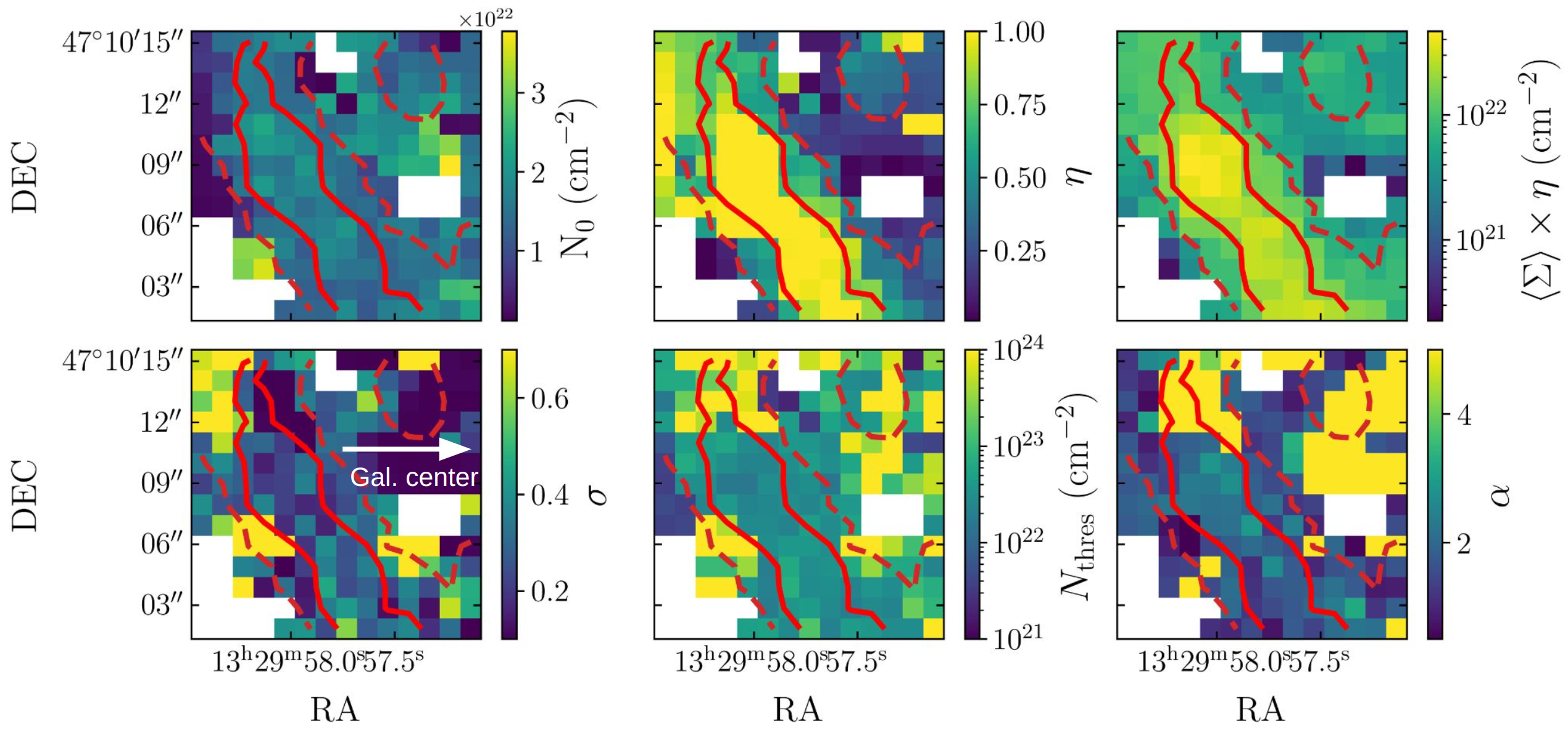}
    \caption{MAP estimations of the sub-beam $N$-PDF parameters across our M51 test region. \textit{Top:} from left to right the panels show the mean column density of the LN part of $N$-PDF ($N_0$), the pixel area filling factor ($\eta$), and the average gas density (including blank sky contributions). \textit{Bottom:} from left to right are displayed the width of the log-normal ($\sigma$), the column density of transition between the log-normal and power-law parts of the $N$-PDF ($N_{\rm{thresh}}$) and the power-law index ($\alpha$). Red contours in each panel indicate \thico{} integrated intensities of 4~and 12~K\,km\,s$^{-1}$ (dashed and solid contours, respectively). The white arrow indicates the direction to the galactic centre. To first order, the gas is denser and more gravitationally unstable inside the spiral arm than outside the arm. 
  }
  \label{fig:m51:param-map}
    \end{figure*}
}

\newcommand{\MdenseMap}{%
    \begin{figure*}
        \centering %
        \includegraphics[width = 0.8\linewidth]{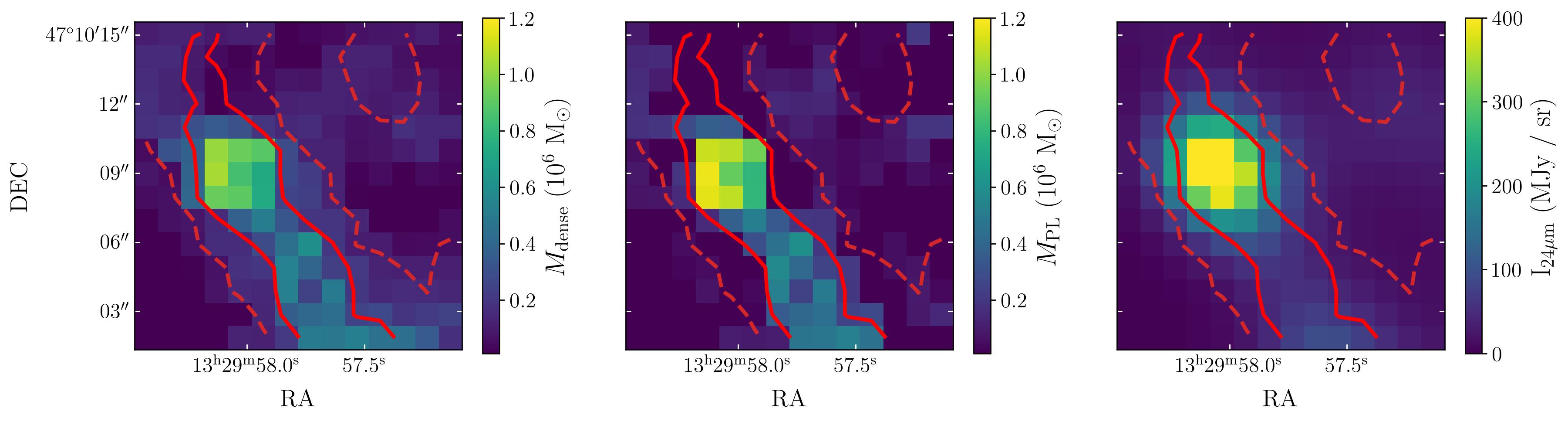}
    \caption{The spatial distribution of
     the mass of dense gas (left), the gas mass in the power-law
     part of the $N$-PDF (middle), and the 24$\,\mu$m surface brightness in our M51 target region. We use the 24$\,\mu$m emission as a proxy for star formation. The masses are derived from the MAP estimate of the $N$-PDF, using equations
     \ref{eq:sigma_dense} and \ref{eq:sigma_pl}. The red contours are the samed as in Figure~\ref{fig:m51:param-map}. The masses of dense and PL gas appear highly correlated, with a similar spatial distribution as the 24$\mu$m emission. } 
    \label{fig:m51:Mdense-map}
    \end{figure*}
}

\newcommand{\MdenseSFR}{%
    \begin{figure}
        \centering %
        \includegraphics[width = 1\linewidth]{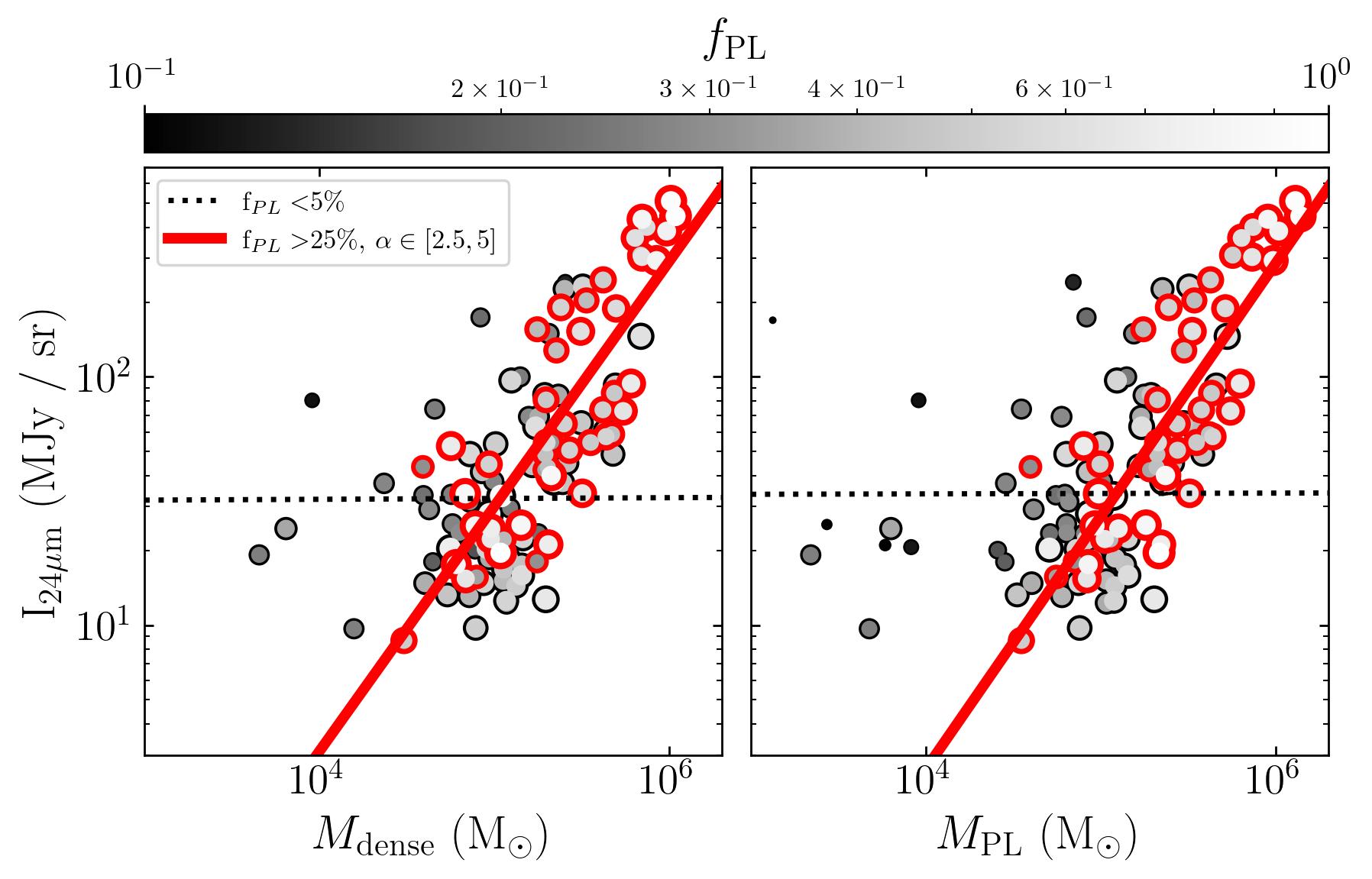} \\

    \caption{Correlation between the 24$\mu$m integrated intensity and the mass of dense gas (\textit{left}) and mass of gas in the power-law part of the $N$-PDF (\textit{right}) for pixels within our M51 test region. Each data point corresponds to a pixel within our field. The symbol size and grey shading represent $f_{\rm PL}$, the mass fraction of the gas in the power-law part of the $N$-PDF. Symbols with a red outline identify pixels where $f_\text{PL} \geq 25\%$ and the slope of the power-law $\alpha \in [2.5,5]$. The dotted line is a linear fit to the pixels where $f_\text{PL} < 5\%$. The thick red line is a fit to the points where $f_\text{PL} \geq 25\%$ and $\alpha \in [2.5,5]$. The latter fit has a correlation coefficient $r = 0.85$ and slope $s = 1.0$.}
    \label{fig:m51:mdense-sfr-corr}
    \end{figure}
}

\newcommand{\FigRatiosNSig}{%
    \begin{figure*}[h]
        \centering %
        \includegraphics[width = 0.9\linewidth]{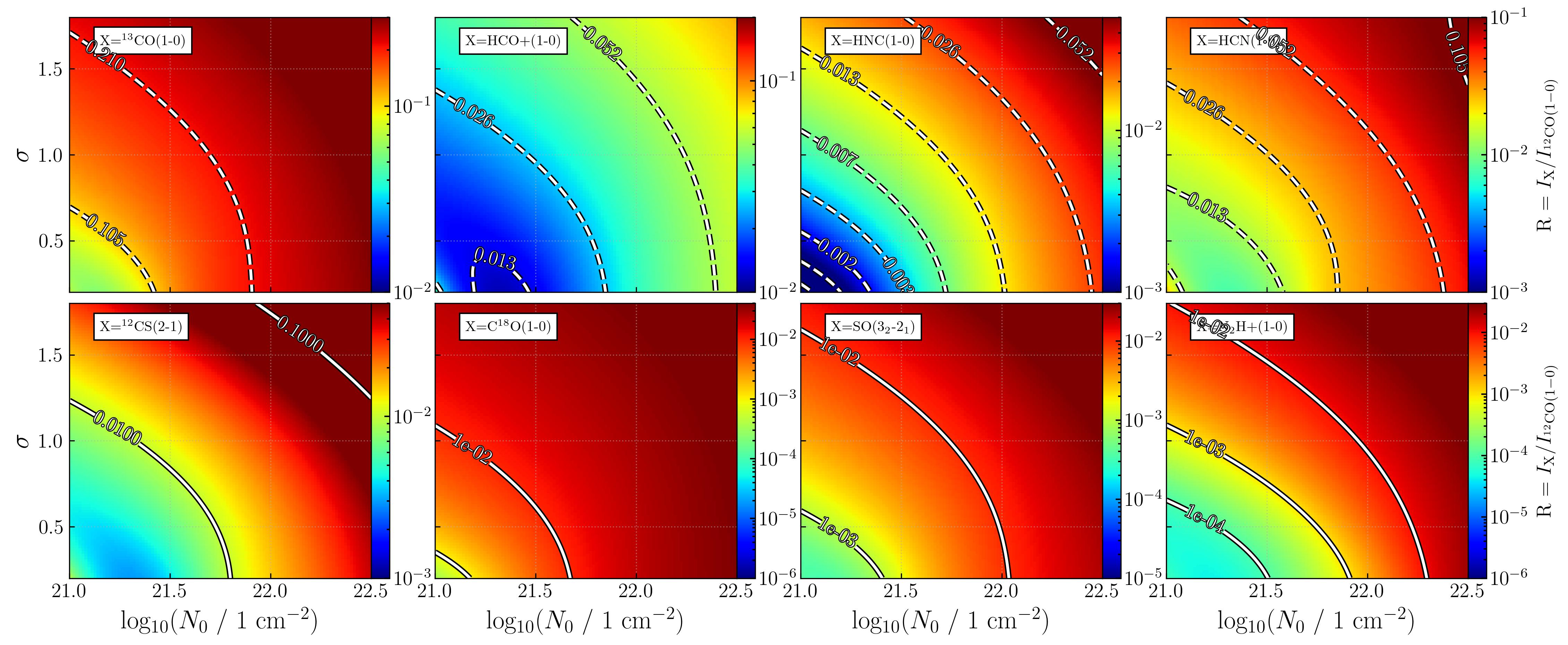}
    \caption{Model predicted line integrated intensities ratios over \tweco{} as a function $\sigma$ and $N_0$ for a purely LN $N$-PDF. Top row shows the ratios of \thico{}, \hcop{}, \hnc{} and \hcn{} over \tweco{}, from left to right. In this top row dashed lines show ratio isocontours increasing by factors of two. Bottom row shows the ratios of \cs{}, \eigco{}, \so{} and \nhp{} over \tweco{}, from left to right. In the bottom row isocontours represent factors of tens.}
    \label{fig:emission-function:N0-sig0}
    \end{figure*}
}

\newcommand{\FigRatiosAlphaSig}{%
    \begin{figure*}[h]
        \centering %
        \includegraphics[width = 0.9\linewidth]{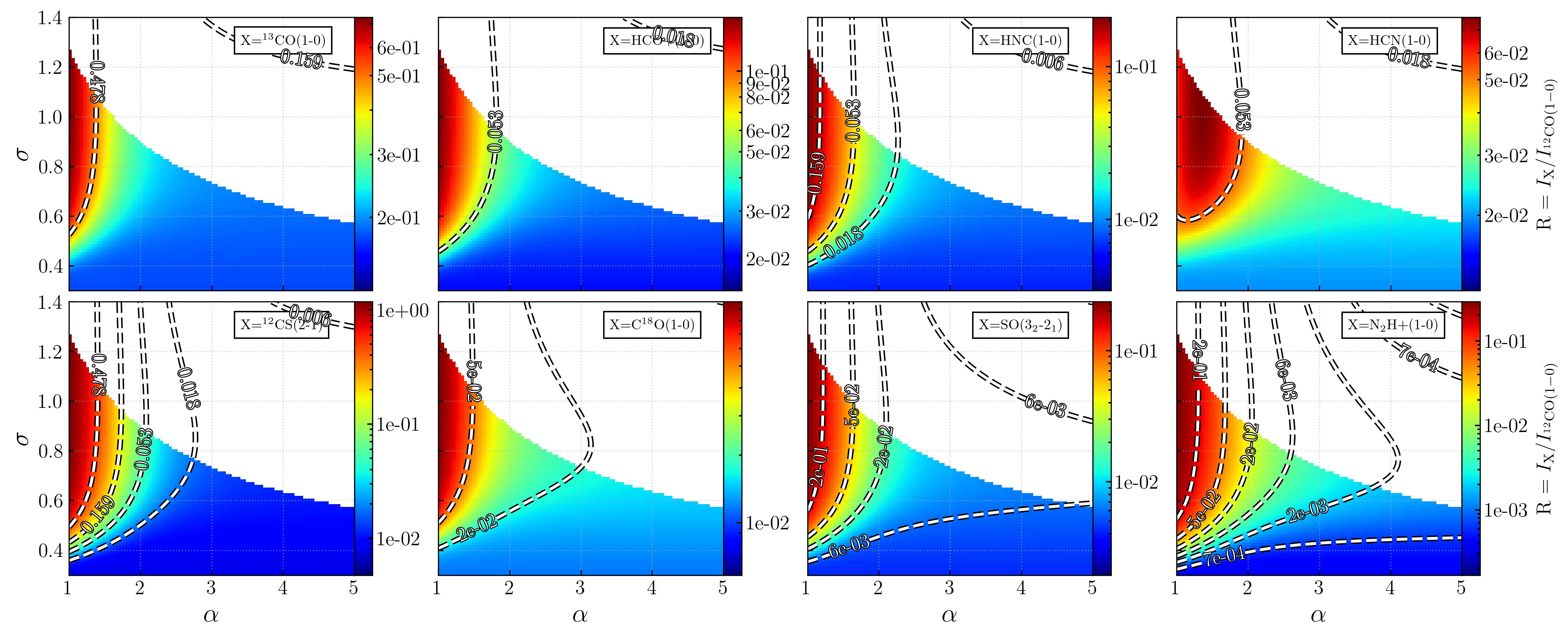}
    \caption{Same as Figure~\ref{fig:emission-function:N0-sig0}, except that the $N$-PDF is composed of a LN and a PL, with varying $\sigma$ and $\alpha$ while $r_{\rm{thres}}$ and $N_0$ are fixed to $N_0 = 5\times10^{22}$\,cm$^{-2}$ and $r_{\rm{thres}}$=5.}
    \label{fig:emission-function:alpha-sig0}
    \end{figure*}
}

\newcommand{\FigRatiosRthresSig}{%
    \begin{figure*}[h]
        \centering %
        \includegraphics[width = 0.9\linewidth]{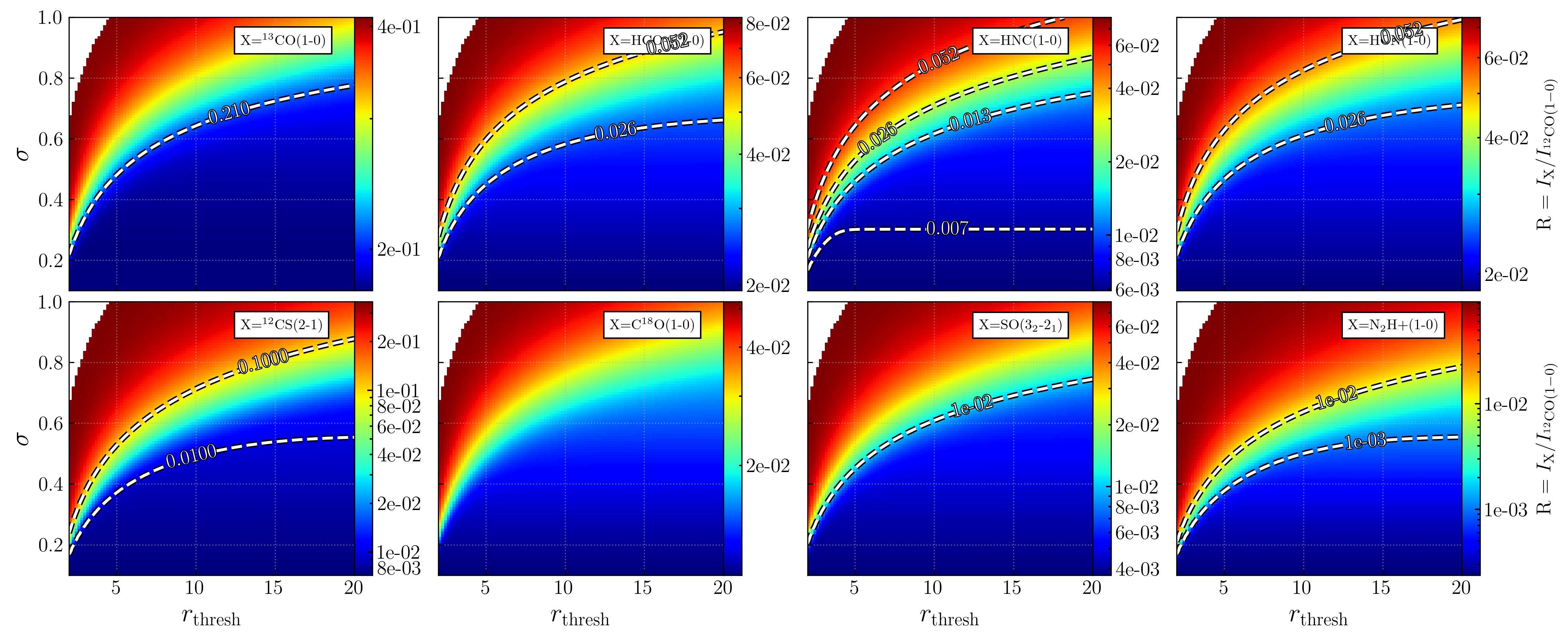}
    \caption{Same as Figure~\ref{fig:emission-function:N0-sig0}, except that the $N$-PDF is composed of a LN and a PL, with varying $r_{\rm{thres}}$ and $\sigma$ while $\alpha$ and $N_0$ are fixed to $N_0 = 5\times10^{22}$\,cm$^{-2}$ and $\alpha=2$.}
    \label{fig:emission-function:rthres-sig0}
    \end{figure*}
}

\newcommand{\FigRatiosRthresAlpha}{%
    \begin{figure*}[h]
        \centering %
        \includegraphics[width = 0.9\linewidth]{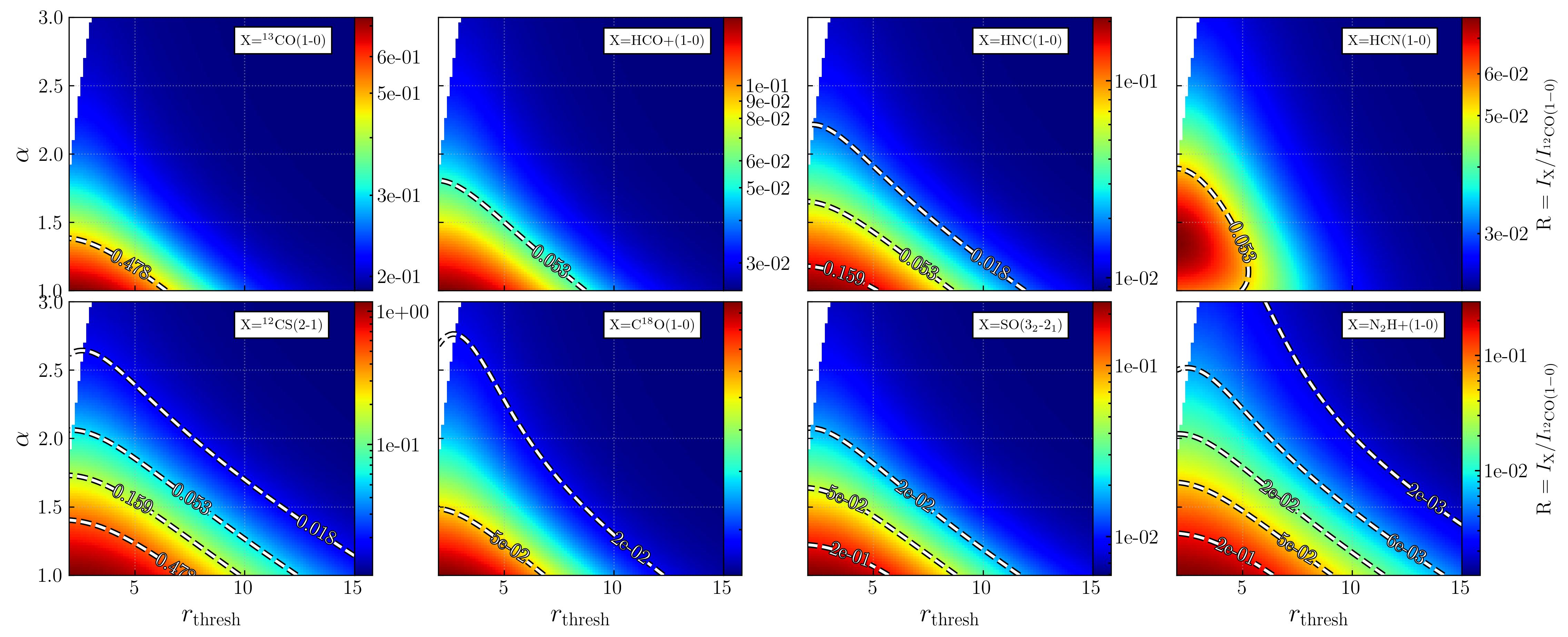}
    \caption{Same as Figure~\ref{fig:emission-function:N0-sig0}, except that the $N$-PDF is composed of a LN and a PL, with varying $r_{\rm{thres}}$ and $\alpha$ while $\sigma$ and $N_0$ are fixed to $N_0 = 5\times10^{22}$\,cm$^{-2}$ and $\sigma=0.6$.}
    \label{fig:emission-function:rthres-alpha}
    \end{figure*}
}

\begin{document}

   \title{Estimating the dense gas mass of molecular clouds using spatially unresolved 3\,mm line observations}

    \author{Antoine Zakardjian\inst{\ref{IRAP}}
          \and Annie Hughes\inst{\ref{IRAP}}
          \and Jérôme Pety\inst{\ref{IRAM},\ref{LERMA/PARIS}} %
          \and Maryvonne Gerin\inst{\ref{LERMA/PARIS}} %
          \and Pierre Palud\inst{\ref{CRISTAL},\ref{LERMA/MEUDON}}
          \and Ivana Beslic\inst{\ref{LERMA/PARIS}} %
          \and Simon Coudé\inst{\ref{WORC},\ref{CfA}} %
          \and Lucas Einig\inst{\ref{IRAM},\ref{GIPSA-Lab}}
          \and Helena Mazurek\inst{\ref{LERMA/PARIS}}
          \and Jan H. Orkisz\inst{\ref{IRAM}} %
          \and Miriam G. Santa-Maria\inst{\ref{UF},\ref{CSIC}} %
          \and Léontine Ségal\inst{\ref{IRAM},\ref{IM2NP}} %
          \and Sophia K. Stuber\inst{\ref{MPIA},\ref{UniHD}}%
          \and S\'ebastien Bardeau\inst{\ref{IRAM}} %
          \and Emeric Bron\inst{\ref{LERMA/MEUDON}} %
          \and Pierre Chainais\inst{\ref{CRISTAL}} %
          \and Karine Demyk\inst{\ref{IRAP}} %
          \and Victor de Souza Magalhaes\inst{\ref{NRAO}}
          \and Javier R. Goicoechea\inst{\ref{CSIC}} %
          \and Pierre Gratier \inst{\ref{LAB}} %
          \and Viviana V. Guzman\inst{\ref{Catholica}} %
          \and David Languignon\inst{\ref{LERMA/MEUDON}} %
          \and François Levrier\inst{\ref{LPENS}} %
          \and Franck Le Petit\inst{\ref{LERMA/MEUDON}} %
          \and Dariusz C. Lis\inst{\ref{JPL}} %
          \and Harvey S. Liszt\inst{\ref{NRAO}} %
          \and Nicolas Peretto\inst{\ref{UC}} %
          \and Antoine Roueff\inst{\ref{IM2NP}} %
          \and Evelyne Roueff\inst{\ref{LERMA/MEUDON}} %
          \and Albrecht Sievers\inst{\ref{IRAM}} %
          \and Pierre-Antoine Thouvenin\inst{\ref{CRISTAL}}
        }

\institute{%
    Institut de Recherche en Astrophysique et Planétologie (IRAP), Université Paul Sabatier, Toulouse cedex 4, France. \label{IRAP} %
    \and
    IRAM, 300 rue de la Piscine, 38406 Saint Martin d'H\`eres,  France. \label{IRAM} %
    \and 
    LUX, Observatoire de Paris, PSL Research University, CNRS, Sorbonne Universit\'es, 75014 Paris, France. \label{LERMA/PARIS} %
    \and 
    Univ. Grenoble Alpes, Inria, CNRS, Grenoble INP, GIPSA-Lab, Grenoble, 38000, France. \label{GIPSA-Lab} %
    \and 
    Univ. Lille, CNRS, Centrale Lille, UMR 9189 - CRIStAL, 59651 Villeneuve d’Ascq, France. \label{CRISTAL} %
    \and 
    LUX, Observatoire de Paris, PSL Research University, CNRS, Sorbonne Universit\'es, 92190 Meudon, France. \label{LERMA/MEUDON} %
    \and
    Department of Earth, Environment, and Physics, Worcester State University, Worcester, MA 01602, USA \label{WORC}
    \and 
    Harvard-Smithsonian Center for Astrophysics, 60 Garden Street, Cambridge, MA, 02138, USA. \label{CfA}
    \and 
    Department of Astronomy, University of Florida, P.O. Box 112055, Gainesville, FL 32611, USA. \label{UF}
    \and
    Instituto de Física Fundamental (CSIC). Calle Serrano 121, 28006, Madrid, Spain. \label{CSIC} %
    \and 
    Université de Toulon, Aix Marseille Univ, CNRS, IM2NP, Toulon, France. \label{IM2NP} %
    \and
    Max-Planck-Institut für Astronomie, Königstuhl 17, 69117 Heidelberg Germany
    \label{MPIA}
    \and
    Fakultät für Physik und Astronomie, Universität Heidelberg, Im Neuenheimer Feld 226, 69120 Heidelberg, Germany
    \label{UniHD}
    \and 
    National Radio Astronomy Observatory, 520 Edgemont Road, Charlottesville, VA, 22903, USA. \label{NRAO} %
    \and
    Department of Physics, University of Connecticut, Storrs, CT, 06269,
    USA \label{UConn}%
    \and 
    Laboratoire d'Astrophysique de Bordeaux, Univ. Bordeaux, CNRS,  B18N, Allee Geoffroy Saint-Hilaire,33615 Pessac, France. \label{LAB} %
    \and 
    Instituto de Astrofísica, Pontificia Universidad Católica de Chile, Av. Vicuña Mackenna 4860, 7820436 Macul, Santiago, Chile. \label{Catholica} %
    \and 
    Laboratoire de Physique de l’Ecole normale supérieure, ENS, Université PSL, CNRS, Sorbonne Université, Université de Paris, Sorbonne Paris Cité, Paris, France. \label{LPENS} %
    \and 
    Jet Propulsion Laboratory, California Institute of Technology,  4800 Oak Grove Drive, Pasadena, CA 91109, USA. \label{JPL}
    \and School of Physics and Astronomy, Cardiff University, Queen's buildings, Cardiff CF24 3AA, UK. \label{UC} %
} 

   \date{Received; accepted}

 
  \abstract
  {Emission lines such as \hcn{} are commonly used by extragalactic studies to trace high density molecular gas (n${_{\rm{H_2}}}>\,\sim10^{4}$ cm$^{-3}$). Recent Milky Way studies have challenged their utility as unambiguous dense gas tracers, suggesting that a large fraction of their emission in nearby clouds is excited in low density gas.}
    {We aim to develop a new method to infer the sub-beam probability density function (PDF) of H$_2$ column densities and the dense gas mass within molecular clouds using spatially unresolved observations of molecular emission lines in the 3\,mm band. }
    {We model spatially unresolved line integrated intensity measurements as the average of an emission function weighted by the sub-beam column density PDF. The emission function, which expresses the line integrated intensity as a function of the gas column density, is an empirical fit to high resolution ($<0.05$\,pc) multi-line observations of the Orion~B molecular cloud. The column density PDF is assumed to be parametric, composed of a lognormal distribution at moderate column densities and a power law distribution at higher column densities. To estimate the sub-beam column density PDF, the emission model is combined with a Bayesian inversion algorithm (implemented in the \textsc{Beetroots} code), which takes account of thermal noise and calibration errors.}
    {
    \modifi{We validate our method by demonstrating that it recovers the true column density PDF of the Orion~B cloud, reproducing the observed emission line integrated intensities within noise and calibration uncertainties. 
    We apply the method to \tweco{}, \thico{}, \eigco{}, \hcn, \hcop{} and \nhp{} observations of a $700\times700$ pc$^2$ field of view (FoV) in the nearby galaxy M51. On average, the model reproduces the observed intensities within 30\%.
    The column density PDFs obtained for the spiral arm region within our test FoV are dominated by a power-law tail at high column densities, with slopes that are consistent with gravitational collapse. Outside the spiral arm, the column density PDFs are predominantly lognormal, consistent with supersonic isothermal turbulence setting the dynamical state of the molecular gas.
    We calculate the mass associated with the power-law tail of the column density PDFs and observe a strong, linear correlation between this mass and the 24$\mu$m surface brightness.}
    }
   {
   \modifi{Our method is a promising approach to infer the physical conditions within extragalactic molecular clouds using spectral line observations that are feasible with current millimetre facilities. Future work will extend the method to include additional physical parameters that are relevant for the dynamical state and star formation activity of molecular clouds.}
   }


   \maketitle
%
\section{Introduction}
\label{sec:intro}


\modifi{Galactic studies show that star formation preferentially
occurs in dense, gravitationally bound substructures within
molecular clouds \citep[e.g.,][and references therein]{Shu1987,Andre2010}. At parsec scales, this substructure is characterised by filaments, spanning several parsecs in length and with a characteristic width of $\sim$0.1\,pc \citep{Arzoumanian2011}. 
Once these filaments reach column densities of $\sim 7 \times 10^{21}$\,cm$^{-2}$, they become gravitationally unstable and tend to fragment into cores \citep{Andre2014}.
The column density threshold for the formation of dense cores corresponds to an observational ''dense gas'' threshold for local clouds, above which the rate of stars formed per dense gas mass is found to be constant \citep{Lada2010}. It has been proposed that the star formation rate (SFR) of a molecular cloud is set by the mass of gas above this threshold, while the mass of dense gas itself is set by the different physical mechanisms of filament formation. This cloud-scale picture of the relationship between the cloud substructure, gas column density distribution and star formation has largely been established from local cloud observations, where the spatial resolution is sufficient to characterise cloud substructures \citep[e.g.,][]{Arzoumanian2011}, the gas column density distribution \citep[e.g.,][]{Schneider2013} and count individual young stellar objects \citep[e.g.,][]{Lada2010}. }

\modifi{Extragalactic observations offer the possibility to study star formation across a wider range of environments and physical conditions than encountered in local clouds. 
Focusing on the properties of dense (n$_{\text{H2}}$ > $\sim$10$^4$ cm$^{-3}$) gas traced by \hcn{} emission, \cite{GaoSolomon2004} showed that there was a linear relationship between the infrared (IR) and \hcn{} luminosities of 65 local galaxies, consistent with a cloud-scale SFR that is proportional to the dense gas mass if the \hcn{} emission is a robust, universal tracer of high density gas.
More recently, \cite{Gallagher2018}, \cite{Jimenez2019} and \cite{Garcia2022}, have highlighted how the correlation between dense gas tracers such as \hcn{} and the star formation rate varies within and among nearby ($d < 20$\,Mpc) galaxies. \cite{Usero2015, Querejeta2019, Beslic2021, Beslic2024} and \cite{Neumann2023} have further shown that the  \hcn{}/\tweco{} ratio, considered to represent the fraction of high density molecular gas relative to the bulk molecular reservoir, varies with galactic environment (centre, bar, spiral arms, disk). Those studies find that the dense gas fraction is often enhanced in the central regions of galaxies, whereas the star formation efficiency in the dense gas (as traced by SFR/\hcn{}) decreases. The dense gas fraction and star formation efficiency have likewise been shown to vary with the cloud-scale molecular gas velocity dispersion and surface density \cite{Gallagher2018b}.}

In practice, extragalactic observations suffer from limited spatial resolution and rely on a subset of bright lines arising from low-J rotational level transitions of molecules with high dipole moments. Such lines are commonly labelled ''dense gas tracers'', referring to their high critical density induced by the high dipole moment of their molecule. The conventional assumption is that \hcn{}, \hcop{} and other dense gas tracers emit predominantly at densities exceeding n${_{\rm{H_2}}}>10^{4}$\,cm$^{-3}$ \citep{GaoSolomon2004}. This ideal notion of the excitation of dense gas tracers has been challenged by recent high resolution observations of nearby Milky Way clouds \citep{Pety2017, Barnes2020, Santa-Maria2023} and simulations of molecular clouds coupled to chemical networks and radiative transfer codes \citep{Priestley2024}. The growing understanding of these tracers is that a significant fraction of their cloud-scale emission can arise from gas at low to moderate densities where the lines are sub-critically excited. \modifi{In this low density regime, for example, \hcn{} sub-thermal excitation resulting from collisions with neutrals and electrons can produce faint emission across a spatially extended region \citep{Goldsmith2017, Goicoechea2022}, since the \hcn{} abundance remains significant in diffuse, UV-illuminated regions \citep{Santa-Maria2023, Liszt2001}. } This more nuanced perspective of the emission from \hcn{} and similar species does not completely undermine their value as gas density tracers in external galaxies. In the absence of massive stars and widespread UV radiation, \hcn{} excitation by electron collisions may be limited, contributing only a small fraction of the total emission measured over large spatial scales. Observationally, \cite{Jimenez2023} have shown that the \hcn{} over \nhp{} ratio remains fairly constant at $\sim0.1 - 1$\,kpc scales in a galactic disk. Since the \nhp{} line has been observed to emit almost exclusively in dense, gravitationally unstable star-forming filaments and cores \cite{Bergin2007, Priestley2024}, a constant \hcn{} over \nhp{} ratio tends to support the validity of \hcn{} as an observational probe of dense gas. In summary, the caveats surrounding the use of \hcn{} and similar lines as dense gas proxies call for better modelling of their spatially unresolved emission, and methods that take into account that extragalactic measurements sample the emission arising from gas with a range of densities that are averaged within a single telescope beam.

\modifi{This paper proposes such a ''beam-averaged'' model, in which the sub-beam distribution of column densities is expressed as a piecewise lognormal (LN) and power-law (PL) distribution \citep[often referred to as a 'gravoturbulent' model in the literature, e.g.,][]{Burkhart2018}. Spatially unresolved integrated intensity is then an average of a sub-beam emission function weighted by the sub-beam column density PDF. 
Other studies have proposed a similar modelling approach in the past \citep[e.g.,]{Leroy2017, Bemis2024}, but were mainly focused on a parametric study of their model. Here, we propose a full Bayesian inversion procedure to retrieve the PDF parameters from unresolved observations. Furthermore, the model presented here relies on an empirical emission function based on the multi-line, high resolution ($<0.01$\,pc) ORION~B survey, to account for both chemical and radiative transfer effects that have an impact on emission lines over several orders of magnitude in column density, whereas previous studies relied on RADEX-based radiative transfer modelling of each resolution element of the sub-beam volume density PDF. Thus, we consider the PDF of column densities (henceforth $N$-PDF) instead of volume density distributions ($n$-PDF).}

The ORION~B survey and other datasets used in this work are described in Section~\ref{sec:data}. The overall ''beam-averaging'' model and the empirical emission function are presented in Section~\ref{sec:model:formulation}. The Bayesian inversion procedure, based on the work of \cite{Palud2023} and the code \textsc{Beetroots}, is summarized in Section~\ref{sec:bayesian-inversion}.
Section~\ref{sec:results-orionb} presents a test of the inversion performance using the spatially averaged ORION~B data, which we also use to explore the model's degeneracies.
In Section~\ref{sec:results-M51}, we apply our inversion method to a 700$\times$700 pc$^2$ (14$\times$14 pixel) field in the nearby galaxy M51, using 3\,mm line observations from the SWAN survey \citep{Stuber2025}.
Limitations and future improvement of the model, comparisons with other ''beam unmixing'' techniques and comparison with the single dense gas tracer and line ratios approaches are discussed in Section~\ref{sec:discussion}. We summarise our key findings and conclusions in Section~\ref{sec:conclusion}. 
\modifr{Illustrations of the current model's most important degeneracies (Sec.~\ref{sec:appendix:degeneracies}), observations compared to model predictions in the Orion~B cloud (Sec.~\ref{sec:appendix:Orionb}), maps of signal-to-noise (SNR) and predicted line intensities in the M51 target region (Sec.~\ref{sec:appendix:M51_obs}) and a parametric study of the model (Sec.~\ref{sec:appendix:parameter-study}) are provided as appendices.}

\section{Data}
\label{sec:data}

\subsection{Orion~B data}


\noindent We use data from the IRAM ORION-B \citep[Outstanding Radio-Imaging of OrioN B, co-PIs: Pety \& Gerin,][]{Pety2017} Large Programme. The observations target a 18$\times$13\,pc region within the Orion~B molecular cloud, a well-known star-forming region at a distance of 410\,pc \citep{Cao2023}. 
ORION-B data surveyed frequencies from 72 to 116.5\,GHz with a spectral resolution of 0.5~km s$^{-1}$. The typical angular resolution of the observations is 25", corresponding to a linear \modifr{resolution (beam size)} of 0.05\,pc, with data gridded on 9" (0.02\,pc) pixels. Depending on the observed frequency, the median noise level in the datacube ranges from 100 to 180\,mK. A more comprehensive description of the ORION-B survey, including a description of the data acquisition and reduction procedures, is presented in \cite{Pety2017}.
In this paper, we use the nine strongest emission lines detected in the ORION-B dataset: \tweco{}, \thico{}, \eigco{}, \hcop{}, \hcn{}, \hnc{}, \cs{}, \so{} and \nhp{}. 



To complete the IRAM~30\,m emission line data, we use column density maps presented by \cite{Lombardi2014}.
These column density maps are derived from dust far-infrared (FIR) and sub-millimetre continuum observations by the \textit{Herschel} Gould Belt Survey \citep{Andre2010, Schneider2013} and Planck satellite \citep{Planck2014}.
Through a fit of the spectral energy distributions constructed from these datasets, \cite{Lombardi2014} inferred the spatial distribution of the dust opacity at 850\,$\mu$\,m ($\tau_{850}$).
 The opacity at 850\,\textmu\,m is converted to a visual extinction via $A_{\mathrm{V}} = 2.7 \times 10^4$ $\tau_{850}$ mag, as described in \cite{Pety2017}. The H$_2$ column density is then estimated from the visual extinction using the conversion factor $N_\mathrm{H_2} / A_{\mathrm{V}} = 0.5 N_\mathrm{HI} / A_{\mathrm{V}} = 0.9 \times 10^{21}$ cm$^{-2}$ mag$^{-1}$.

\subsection{M51 data}


We use observations of molecular line emission in M51 \citep[distance = 8.58\,Mpc,][]{McQuinn2016} from the SWAN \citep{Stuber2025} and PAWS \citep{Schinnerer2013} surveys. SWAN observed the \thico{}, \eigco{}, \hcop{}, \hcn{}, \hnc{} and \nhp{} lines at an angular resolution of 3" and a spectral resolution of 10\,km\,s$^{-1}$ across the central $5\times7$\,kpc$^2$ part of M51.
PAWS provides complementary \tweco{} observations, which were spatially and spectrally smoothed to match the resolution of the SWAN data. 


We use publicly available \textit{Spitzer} 24$\mu$m observations of M51 as a proxy of embedded star formation activity. The \textit{Spitzer} map that we use is presented in \cite{Dumas2011}. The observations have been deconvolved with the \textit{HiRes} algorithm \citep{Backus2005} in order to achieve an angular resolution of 3" \modifr{(140\,pc)}.  

\section{Modelling spatially unresolved emission}
\label{sec:model}

\subsection {Formulation of the problem}
\label{sec:model:formulation}

For spatially unresolved emission lines, the observed integrated intensity is an average of the sub-beam distribution of integrated intensities arising from the two-dimensional projection of the molecular gas onto the plane of the sky. \modifi{Given a function $f(\boldsymbol{\theta}): \mathbb{R^D} \rightarrow \mathbb{R}^L$ relating the sub-beam intensities to the sub-beam physical parameters $\boldsymbol{\theta} \in \mathbb{R}^D$, the relation between the observed integrated intensities of a set of $L$ emission lines $\textbf{y} = \{y_{1},...,y_{\ell}, ...,y_{L}\} \in \mathbb{R}^L$ and the sub-beam density distribution of physical parameters $p(\boldsymbol{\theta})$ can be expressed as:}

\begin{ceqn}
\begin{align}\label{eq:mixing:all}
        \textbf{y} = \modifr{\eta} \, \int f(\boldsymbol{\theta})\,p(\boldsymbol{\theta})\,\diff{}\boldsymbol{\theta}\modifr{,}
\end{align}
\end{ceqn}
\modifr{where $\eta$ is the pixel area filling factor of the molecular gas, accounting for intensity dilution due to blank sky contributions if molecular emission is present only in a fraction of the pixel area.}
Here, the function $f(\theta)$ is known and can be highly complex, for instance a sophisticated model including chemistry, realistic cloud geometry and radiative transfer. \modifrr{The distribution $p(\theta)$ integrates to one by definition.}

\modifi{In general, this equation is impossible to invert, since the integration over an unknown multi-dimensional distribution makes it highly degenerate. To make the model invertible, the function $f(\boldsymbol{\theta})$ should be as simple as possible, focusing on the most sensitive variable $\theta$.
In this paper, the physical parameter of interest is $N_{\mathrm{H_2}}$, which is also the parameter that predominantly drives the intensity of molecular emission lines \citep{Gratier2017, Gratier2021}. Knowing this, the problem can be greatly simplified by reducing the parameter space to $N_{\mathrm{H_2}}$ only:}

\begin{ceqn}
\begin{align}
        \textbf{y} = \modifr{\eta} \int f(N_{\mathrm{H_2}})\,p(N_{\mathrm{H_2}})\,\diff{} N_{\mathrm{H_2}},
\end{align}
\label{eq:mixing:NH2}
\end{ceqn}

\noindent The contribution of other parameters to the emission is either considered to be negligible, or encoded in $f(N_{\mathrm{H_2}})$. The main difficulty is to retrieve the continuous distribution $p_{N_{\mathrm{H_2}}}$ from a small \modifi{($L$<10)} set of observations. This can be achieved by either discretizing the distribution into bins, or assuming that the distribution can be parametrized with a small set of parameters $\phi$.
\modifi{When discretising the distribution, the free parameters $\phi$ are the height of the individual bins, thus limiting the column density resolution that can be achieved to the number of independent observations.}
Since extragalactic millimetre surveys only target a handful of emission lines, \modifi{we prefer to assume a parameterised form for $p(N_{\mathrm{H_2}})$, ensuring that we have fewer parameters than independent observations} and adopt physically motivated assumptions for the shape of the column density PDF.

\subsection {Parametrization of the gas column density distribution}
\label{sec:model:parametrisation}


\modifi{To select the appropriate parametric family of distributions representing $N$-PDFs, we examine the shapes of $N$-PDFs determined from both numerical simulations and observations.
In this Section, we discuss how the shape of the column density PDF depends on turbulence and gravity, the dominant physical processes that, together with the magnetic field, determine the dynamical state of the molecular gas \cite{MacLowKlessen2004}.}


\subsubsection{Isothermal supersonic turbulence}
\label{sec:model:parametrisation:LN}

The volume and column density PDF of molecular and atomic gas depends largely on turbulence.
Two common assumptions about turbulence in cold and dense ISM are that the gas is isothermal and the turbulent motions are supersonic \cite{Larson1981}.
An analytical prediction and consequence of supersonic isothermal turbulence is the parametrization of the density PDF (whether volume or column density) as a lognormal (LN) \citep{Vazquez-Semadeni1994}. 
This type of distribution has been observed in numerical simulations \citep[e.g.,][]{Nordlund1999, Federrath2008} and confirmed by observations \citep[e.g.,][]{Kainulainen2009}. 
Qualitatively, a LN distribution naturally arises in a multiplicative stochastic process. Density fluctuations may be regarded as the result of a product of independent and identically distributed random variables (successive shock compressions). The log of these density variations then turns into a sum of random variables, which approaches a normal distribution under the central limit theorem.
\modifi{The column density distribution with mean $N_0$ and standard deviation $\sigma$ can then be expressed as:}

\begin{ceqn}
\begin{align}\label{eq:LN}
   p(N_{\rm{H_2}}) = \frac{1}{ N_{\mathrm{H_2}} / N_0 \sqrt{2 \pi \sigma^2} } \exp \left(- \frac{(\ln(N_{\mathrm{H_2}} / N_0) - \mu)^2}{2 \sigma^2}\right),
\end{align}
\end{ceqn}
\modifi{where the expectation $\mu$ is related to the variance $\sigma$ via $\mu = - \sigma^2 / 2$. This condition is mostly relevant for numerical simulation and imposed by mass conservation requirements in the closed box simulation \citep{Nordlund1999}.
While $\mu=0$ could be used \citep[e.g.,][]{Vazquez-Semadeni1994} here, we still chose to impose the above relation between mean and variance in order to facilitate the comparison with numerical simulations. }





\subsubsection{Gravitational collapse}
\label{sec:model:parametrisation:PL}

Turbulence is not the sole mechanism affecting the gas density distribution in molecular clouds. Gravity, as it becomes dominant over turbulence at higher gas densities, will tend to produce a power-law (PL) tail with index $-\alpha$ such that :

\begin{ceqn}
\begin{align}\label{eq:PL_N_obs}
p(N_{\mathrm{H_2}}) = p_0 \, {N_{\mathrm{H_2}}}^{-\alpha}\modifr{.}
\end{align}
\end{ceqn}

The emergence of a PL tail in volume and column density PDFs was first observed in numerical simulations \citep[e.g.,][]{Slyz2005, Kritsuk2011, Federrath2013} and later detected in nearby molecular clouds \citep{Kainulainen2009, Schneider2022}.  Such PL tails are generally attributed to gravitationally collapsing gas. Indeed, a collapsing homogeneous sphere of gas is expected to develop a radial profile of density $\rho(r) = r^{-k}$ \citep{Shu1977}.  As shown by \cite{Federrath2013}, in this scenario the volume density PDF defined as $p(\rho) \propto dV / d\rho$ becomes:

\begin{ceqn}
\begin{align}
    p(\rho) \propto \frac{dV}{d\rho} \propto \frac{dV}{dr}\frac{dr}{d\rho} \propto \rho^{-3/k-1},
\end{align}
\end{ceqn}
\modifr{with $V \propto r^3$ the volume}. For column densities, the radial density profile is $N(r) \propto \rho r \propto r^{-k+1}$ and the distribution of column density becomes:

\begin{ceqn}
\begin{align}\label{eq:PL_N_theo}
    p(N) \propto \frac{dA}{\modifr{dN}} \propto N^\frac{1+k}{1-k},
\end{align}
\end{ceqn}
\modifr{with $A\propto r^2$} the area. Relating equation \ref{eq:PL_N_obs} and \ref{eq:PL_N_theo}, the PL index $\alpha$ of the $N$-PDF can therefore be associated to an equivalent radial density profile of \modifi{index $k = (1 + \alpha)/(\alpha-1)$.}
The analytical expectation for $k$ in a collapsing isothermal sphere is 2 \modifi{(corresponding to $\alpha = 3$)} in the outer static envelope of the collapsing sphere and 3/2 \modifi{($\alpha = 5$)} in the free-falling inner envelope \citep{Shu1977}.


\cite{Arzoumanian2011} estimated the radial density profile of filamentary structures in the IC 5146 molecular cloud using \textit{Herschel} observations. They reported indices $k \in [1.5 : 2.5]$, equivalent to $\alpha = 2.3 - 5$.
In numerical simulations, \cite{Federrath2013} observed PL tails at higher densities, with values of $k$ also ranging from 1.5 to 2.5.
\modifi{Across Galactic molecular clouds, \cite{Schneider2022} observed PLs with slopes $\alpha$ ranging from 2 to 5.}


\subsection {Empirical emission function}
\label{sec:model:emission-function}

\FigEmissionFunction{}

In this paper, we use an empirically motivated emission function $W\,(\text{K km s$^{-1}$) = }f(N_{H_2})$ that we obtain by fitting the ORION-B data.  This choice is motivated by the need to capture not only the full complexity of emission lines' excitation mechanisms, but also the chemical processes altering molecular abundances. Indeed, the emission function must include several orders of magnitude in column density, covering both diffuse, illuminated regions and dense, well-shielded cores. The chemistry, and in particular photodissociation and molecular freeze out onto dust grains, must be accounted for, which is not the case for emission functions developed solely on radiative transfer calculations.

The choice of an empirical function is also partially motivated by recent observations of Milky Way clouds \citep{Pety2017, Barnes2020, Tafalla2021, Tafalla2023} and numerical simulations \citep{Priestley2024}, which display similar emission functions even though the observed clouds harbour a wide range of star formation activity. 
The seemingly uniform behaviour of molecular line integrated intensities as a function of column density across nearby molecular clouds suggests that Galactic observations could be used to generate a calibrated emission function for extragalactic applications.
\modifr{Among available datasets, ORION~B provides the best combination of high spatial resolution, sensitivity, and broad frequency coverage over a cloud-scale field of view, hence its use as the basis for our empirical emission function.}

In order to fit the relation between line integrated intensities and column densities, we bin the ORION~B data in equally-sized logarithmic bins of column density. In practice, we split the data in 30 bins, which we find to be a good compromise for establishing the average trend without excessively smoothing the variations. \modifr{Figure \ref{fig:emission-function}} shows the resulting binned trend between column density and the integrated intensities of the emission lines that we include in our model, as well as the standard deviation in each bin. 

For each emission line, the trend resembles a combination of two power laws with a smooth transition, \modifi{which we fit using $\chi^2$ minimization of the following function:}

\begin{equation}
    f_{\ell}(N_{\rm{H_2}}) = f_{l,b}\,\left( \frac{N_{\rm{H_2}}}{N_b} \right)^{-\beta_1} \left\{ \frac{1}{2} \left[ 1 + \left( \frac{N_{\rm{H_2}}}{N_b} \right)^{\frac{1}{\Delta}} \right] \right\}^{(\beta_1 - \beta_2) \Delta}.
\label{eq:bkn_pl}
\end{equation}

In this equation, $N_b$ is the break location, $\beta_1$ and $\beta_2$ are the slopes before and after the break, $\Delta$ is the smoothing parameter of the breaking point \modifr{and $f_{l,b}$ is the function value at the break point}. Bins at lower column density with little or no data are discarded. The functions $f_l$ are specific to each of the $l$ emission lines, and their fit parameters are summarized in Table~\ref{tab:bkn_pl_params}.

The scatter in the trends depends on the emission line considered. \tweco{} and \thico{} show little scatter, whereas \hcn{}, \hcop{} or \cs{} display significantly larger dispersion. The dispersion around the mean trend increases with decreasing column density, which can be explained by the averaging of environments with varying radiation field. Indeed, the ORION~B field of view displays a clear gradient of $G_0$ from east to the west. Bins of lower column density corresponding to the outermost layers of the cloud can thus include diffuse and translucent regions with a wide range of UV-illumination. 

A noticeable feature of these trends is the intensity of the \tweco{} line. It is significantly higher than the values observed by \cite{Tafalla2021,Tafalla2023} in nearby clouds. 
The enhanced \tweco{} emission may be due to a higher gas temperature, driven by the high FUV radiation field pervading Orion~B \citep{Pety2017, Santa-Maria2023}. 

Besides radiative transfer and collisional excitation of the lines, \cite{Tafalla2023} explained the similarity of the trends they observed by transitions through different chemical regimes with increasing visual extinction.  
At column densities $N_{\rm{H_2}} \leq 1-2 \times 10^{21}$ cm$^{-2}$, the emission of CO molecules decreases sharply as the molecules in the surface layer of the cloud are photodissociated by the external radiation field. 
Above this column density threshold, shielding mostly prevents photodissociation and abundances remain relatively constant.
The column density \NH = $1-2 \times 10^{22}$ cm$^{-2}$ marks the transition to the ''freeze-out'' regime, where carbon molecules start to freeze onto dust grains. As a result, the abundances of C-bearing species steadily decrease as they are removed from the gas phase.
Finally, column densities above \NH = $10^{23}$ cm$^{-2}$ are associated with regions dominated by stellar feedback. This regime sees increased emission from most species, in particular \hcn{} and \cs{}, which is attributed to shocks and a substantial temperature increase affecting the chemistry of these molecules. The trends between line emission and column density described by \cite{Tafalla2023} are similar to what we observed in Orion~B. 

In conclusion, we adopt a realistic, albeit rigid, empirical emission function, which we construct using a fit to the ORION~B data. This implies that the method is applicable to regions where the metallicity, temperature, radiation and abundance are relatively similar to conditions in Orion~B. Generalizing the emission function  will be the subject of future work.
We outline some caveats of the emission function and potential avenues to increase its flexibility in Section~\ref{sec:discussion:limitation-function}.

\subsection{Global beam-mixing model}
\label{sec:model:global-model}

The final emission model reads :

\begin{ceqn}
\begin{align}\label{eq:final_model}
        y_{\ell} = g_{\ell}(\phi) = \modifr{\eta} \int f_{\ell}(N_{\mathrm{H_2}})\,p_{\phi}(N_{\mathrm{H_2}})\,\diff{} N_{\mathrm{H_2}},
\end{align}
\end{ceqn}
with $f_{\ell}(N_{\mathrm{H_2}})$ given by equation \ref{eq:bkn_pl} and the parameters listed in Table~\ref{tab:bkn_pl_params},
while $p_{\phi}(N_{\mathrm{H_2}})$ is defined as :

\begin{ceqn}
\begin{align}
p_{\phi}(N_{\mathrm{H_2}}) \, \modifrrr{=C\times}
\begin{cases}\label{eq:final_pdf}
\frac{1}{ N_{\mathrm{H_2}} / N_0 \sqrt{2 \pi \sigma^2} } \exp \left(- \frac{(\ln(N_{\mathrm{H_2}} / N_0) - \mu)^2}{2 \sigma^2}\right)&\text{, if } N_{\mathrm{H_2}}<N_\text{thresh}, \\
    {p_0 \, N_{\mathrm{H_2}}}^{-\alpha}&\text{, if } N_{\mathrm{H_2}} \geq N_\text{thresh},
\end{cases}
\end{align}
\end{ceqn}
with \modifrrr{$C$ a normalisation constant and} parameter $p_0$ a constant ensuring continuity between the LN and PL segments. Its value is given by relating equations \ref{eq:LN} and \ref{eq:PL_N_obs} at $N_{\mathrm{H_2}} = N_{\rm{thresh}}$. In practice, we express $N_{\rm{thresh}}$ as a function of $N_{0}$ and $\mu$ as $N_{\rm{thresh}} = r_{\rm thresh} N_{0} \exp(\mu)$, to ensure that $N_{\rm{thresh}} < N_{0} \exp(\mu)$. \modifr{Similarly, we express $\alpha$ as a function of the parameter $d_\alpha = -\alpha / \left[ \frac{\diff{} \log( p{_{\rm{LN}}}(N_{\rm{H_2}}))}{\diff{} \log(N_{\rm{H_2}})}|_{N_\mathrm{thresh}} \right] < 1$. This condition ensures that the PL tail is not steeper than the LN beyond the transition density $N_{\rm{thres}}$}
\modifr{In total, our model has five parameters $\phi = [N_{0}, \sigma, r_{\rm thresh}, d_\alpha, \eta]$}.
As there is no closed-form expression to Equation~\ref{eq:final_model}, the integral is computed numerically using the trapezoidal rule over a logarithmically spaced array of 100 column densities ranging from 10$^{20}$ to 10$^{24}$\,cm$^{-2}$. The \modifrrr{constant $C$ in Equation~\ref{eq:final_pdf} is also computed numerically so that the $N$-PDF integrates to one.}

The LN width $\sigma$, the relative transition density $r_{\rm{thresh}}$ and the PL index $\alpha$ could also be related as $r_{\rm{thresh}} = (\alpha - 1)\sigma^2$ if we impose a seamless transition between the LN and PL regime. This formulation removes one free parameter from the model. Along these lines, \cite{Burkhart2018, Burkhart2019} expressed the transition (volume) density as a function of $\alpha$ and $\sigma$, highlighting the transition density as the critical density for star formation and its relationship to the post-shock critical density for collapse (i.e., density at which the Jeans length is comparable to the sonic length). However, the $N$-PDFs of some local clouds display non-continuous LN to PL transitions. To give our model more flexibility, we thus chose to keep the transition density parameter free, which is why we refrain from referring to our LN + PL model as a 'gravoturbulent' model.

\modifrrr{It should be noted that the filling factor $\eta$ introduced in this model is different from the standard line beam filling factor $\eta_{X}$.
With our modelling approach, $\eta_{X} = \eta \times \int_{X}^{+\infty} p_\phi(N_{H_2}) \, dN_{H_2}$, where $X$ refers to the smallest column density for which the line is detected in the ORION~B dataset.
In other words, $\eta$ is the filling factor of the gas whose column density is above \NH{}$>10^{21}$\,cm$^{-2}$, whereas $\eta_X$ is the filling factor of the considered line.
In the ORION~B dataset, the 1-$\sigma$ sensitivity for the \tweco{} line is 0.3\,K\,km\,s$^{-1}$ per pixel, which corresponds to a column density of $10^{21}$\,cm$^{-2}$ (see Figure~\ref{fig:emission-function}). In this specific case, $\eta = \eta_{\rm{^{12}CO}(1-0)}$.}

\subsection{Average densities and dense gas fractions}
\label{sec:model:parameters}

Given the column density PDF \modifr{$p_{\phi}(N_{\mathrm{H_{2}}})$}, a number of relevant molecular gas properties can be computed.
The mass-weighted average surface density can be computed as:
\begin{equation}
        \langle \Sigma \rangle \,  = \int_{{-\infty}}^{+\infty} N_{H_2} \, p_\phi(N_{H_2}) \, dN_{H_2}\modifr{.} \\[1em]    
\label{eq:sigma}
\end{equation}

The mass-weighted average surface density of gas in the PL regime is:
\begin{equation}
        \langle \Sigma_{\text{PL}} \rangle \,  = \int_{N_{\text{thresh}}}^{+\infty} N_{H_2} \, p_\phi(N_{H_2}) \, dN_{H_2}\modifr{.} \\[1em]    
\label{eq:sigma_pl}
\end{equation}

Finally, the mass-weighted average surface density of the ''dense'' gas is similarly defined as:
\begin{equation}
        \langle \Sigma_\text{dense} \rangle\, = \int_{10^{22}\,cm^{-2}}^{+\infty} N_{H_2} \, p_\phi(N_{H_2}) \, dN_{H_2}, \\[1em]    
\label{eq:sigma_dense}
\end{equation}
where ''dense'' refers to molecular gas at column densities above $10^{22}$\,cm$^{-2}$ to be consistent with Milky Way studies \citep{Lada2010}.
\modifrr{These surface density quantities directly depend on the value of $N_0$, which may be degenerated with the value of $\eta$ (see Appendix~\ref{sec:appendix:degeneracies}).}

\modifrr{We also define quantities which are the product of surface densities with the filling factor $\eta$, and are thus unaffected by the potential degeneracy between $N_0$ and $\eta$: total, dense and power-law gas masses.
The fractions of dense and PL gas can then be computed as $f_\text{dense} = \langle \Sigma_\text{dense} \rangle / \langle \Sigma \rangle$ and $f_\text{PL} = \langle \Sigma_\text{PL} \rangle / \langle \Sigma \rangle$. Similarly, these quantities are unaffected by potential $N_0$-$\eta$ degeneracies.}

\subsection{\modifr{Comparison with other unmixing approaches}}

\label{sec:discussion:comparison-unmixing}

\modifj{Unmixing
     sub-beam physical and chemical parameters is a long-standing problem
     in astronomy and ISM studies.}
Several approaches have been developed to tackle this problem. 

The most widely used model-free methods are inspired by or adapted from Earth observations methods of hyperspectral unmixing. The most commonly used of these methods is Non-negative Matrix Factorization, which blindly decomposes multi-spectral or hyperspectral observations into a linear combination of discrete independent components (referred to as end-members in the hyperspectral unmixing literature). These components can then \textit{a posteriori} be attributed to specific environments, such as different dust grain populations \citep[e.g.][]{Berne2007, de_Mijolla2024, Kishikawa2024}.  

Most methods nevertheless rely on physical models, whether numerical, analytical or empirical. These methods can be subdivided into two categories, depending on whether they assume a discrete or parametric sub-beam distribution of the physical parameters. For the discrete case, the methods used usually rely on one or two components (or zones) \citep[e.g.,][]{Kaneko2023,  Lizée2022, Ramambason2020, Vollmer2017}. One-zone LTE or non-LTE modelling using a single set of average parameters \citep[e.g.,][]{Roueff2021, Roueff2024} would fit into this category. Methods relying on a high number of components have been developed, components which can even be divided in different sub-components \citep[e.g.,][]{Ramambason2022}. The alternative to the discrete case is to perform a parametric estimation of the  sub-beam distribution of parameters. This requires prior knowledge on the shape of the distribution (characterised by a limited set of parameters) to be estimated. It is able to retrieve a complete, continuous distributions of physical properties with a minimum number of model parameters \citep[e.g.,][]{Ramambason2024, Varese2024, Villa_Velez2024}. The method presented here falls into this last category.

In Earth observations, ''unmixing'' traditionally refers to disentangling spatially-averaged emission arising from specific environments. In astronomy, however, the mixing is three-dimensional. Emission mixing along the line-of-sight becomes increasingly important compared to spatial averaging as the linear resolution of observations increases, to the point where the linear beam size is negligible relative to the observed gas length along the line-of-sight. Several methods have been specifically developed to perform line-of-sight unmixing. \cite{Segal2024}, for example, have recently proposed a discrete approach, modelling the gas structure using a three zone 'sandwich' model, that is two outer layers (a foreground and a background layer) surrounding an inner layer, with parameters to be determined in each layer. Our method could be extended to incorporate line-of-sight mixing by adapting Equation~\ref{eq:mixing:all} to observables that are optically thin. In this case, our method could be used estimate the PDF of other parameters (e.g., the radiation field intensity $G_0$) along sightlines of high linear resolution observations, once an appropriate emission function had been established.

\section {Beam unmixing using Bayesian inversion}
\label{sec:bayesian-inversion}

\subsection{The Bayesian Framework}
\label{sec:bayesian-inversion:framework}



Estimating the physical parameters of a system given a model and a
set of observations is a common problem in astrophysics. Our
problem is notably degenerate, since the observables arise from a sum
over a PDF.  Moreover, most of our observations have low to modest
signal-to-noise ratios. We choose to solve it through a Bayesian
framework to retrieve the distribution of solutions in the parameter
space that is compatible with the observations, within noise and
calibration uncertainties. Consequently, it can identify
potential degeneracies in the solutions.  The distribution of
solutions $\pi(\Theta \mid Y)$ is called the ''posterior distribution'', and is
expressed by the Bayes' theorem:

\begin{equation}
\pi(\Theta \mid \text{Y}) = \frac{\pi(\text{Y} \mid \Theta) \pi(\Theta)}{\pi(\text{Y})} \propto \pi(\text{Y} \mid \Theta) \pi(\Theta).
\label{eq:bayes}
\end{equation}
The ''prior distribution'' $\pi(\Theta)$ describes all \textit{a priori} knowledge about the physical parameters that are to be estimated. This prior is coupled with the likelihood function $\pi(Y \mid \Theta)$, which gains more weight as information is provided, to produce the posterior distribution $\pi(\Theta \mid Y)$. The distribution $\pi(Y)$ is a normalization constant, the Bayesian evidence, which is not necessary to compute in Bayesian inference, hence the simplification of equation \ref{eq:bayes}.
Once sampled, the posterior distribution can be used to compute different estimators of the ''best'' physical parameters describing the observations.


We use the method and the associated \textsc{Beetroots}\footnote{\url{https://beetroots.readthedocs.io/en/latest/}} code, described
in detail by \cite{Palud2023}. This method was specifically designed
to retrieve model parameters from emission line observations, with a
specific observational model including both noise and calibration
uncertainties, and a sampling algorithm that efficiently samples the
posterior distribution. Here, we summarize the main elements of the
method, namely the noise model, prior information, sampling algorithm and
estimators.

\subsection{Observation and noise model}
\label{sec:bayesian-inversion:noise-model}

The noise model is a critical part of the Bayesian framework that describes how the different sources of noise degrade the observations, for instance through thermal noise and calibration errors.
The observation model used in this work is, for a given pixel $n$:

\begin{equation}
    y_{n\ell} = \varepsilon_{n\ell}^{(m)} g_{\ell}({\phi}_n) + \varepsilon_{n\ell}^{(a)}.
\end{equation}

Here, $\varepsilon_{n,\ell}^{(m)} \sim \log \mathcal{N}\left(-\frac{\sigma_m^2}{2}, \sigma_m^2\right)$ is a LN multiplicative noise expressing calibration errors, and $ \varepsilon_{n,\ell}^{(a)} \sim \mathcal{N}(0, \sigma_a^2)$ is an additive Gaussian white noise associated to thermal noise. 
In this model, $\varepsilon_{n,\ell}^{(m)}$ and $\varepsilon_{n,\ell}^{(a)}$ are assumed to be independent and their variances $\sigma_m$ and $\sigma_a$ known. 

\subsection{Prior information}
\label{sec:bayesian-inversion:prior-info}

The prior distribution $\pi(\Theta)$ is another key element of the Bayesian inversion framework. This distribution encapsulates the prior information about the physical parameters to be estimated. In our case, there is no strong assumption about the parameters, aside from an expected range of acceptable values. Therefore we use a uniform prior distribution on $\Theta$ \citep{Jeffreys1946}, with the validity intervals listed in Table~\ref{tab:val_intervals}.
Given that all parameters range several orders of magnitude, their distribution is set to be uniform in log scale.


\subsection{Sampling}
\label{sec:bayesian-inversion:sampling}

In the case where the posterior distribution has no analytical formulation, Monte Carlo methods become necessary. In Bayesian inference, Markov Chain Monte Carlo (MCMC) algorithms are the most widely employed Monte Carlo technique.

The model presented here is non-linear, \modifi{which implies potential local minima, and must be inverted over large ($>10\,000$ pixel) maps. These challenges are met by the specific MCMC algorithm implemented in \textsc{Beetroots}.} This algorithm combines two complementary sampling kernels: the multiple-try Metropolis kernel (MTM) and the preconditioned Metropolis-adjusted Langevin algorithm (P-MALA). The MTM kernel facilitates a global exploration of the parameter space and escape from local minima, while the P-MALA samples the posterior PDF in and around the local minima.
The combined ''global'' and ''local'' exploration of these kernels allow to efficiently sample the posterior PDF. A complete description of the MTM and P-MALA kernels used in \textsc{Beetroots} is presented in \cite{Palud2023}.

\subsection{Estimators and uncertainty intervals}
\label{sec:bayesian-inversion:estimators}

Different estimators of the parameters can be calculated from the posterior distribution, each with their own merits and drawbacks. Here we use the Maximum A Posteriori (MAP) estimator, which maximizes the posterior distribution: 

\begin{equation}
    \widehat{\Theta}_{\text{MAP}}  = \underset{\Theta}{\arg\min} \, \left[ -\log \pi(Y \mid \Theta) - \log \pi(\Theta) \right].
\end{equation}
\modifi{We prefer it to the more commonly used Minimum Mean Square Error (MMSE), which is given by the mean of the posterior distribution, as degeneracies in the model can lead to strongly skewed posterior PDFs.}

Uncertainty intervals on the estimated parameters can be derived straightforwardly from the posterior PDF. Here, we use the 16\%-84\% percentile range on a parameter's posterior PDF to quantify its uncertainty.

\section{Testing on the ORION~B dataset}
\label{sec:results-orionb}

\PostPDFExample{}
\PostPDFExampleCorr{}

As an initial test, we apply the method to the ORION-B dataset itself.  We spatially and spectrally average the entire datacubes (including noise pixels) to mimic unresolved line observations of
a giant molecular cloud. The resulting vector of averaged intensities is
then used to infer the $N$-PDF. We use the dust column density map derived by \cite{Lombardi2014} to compute the ''true'' $N$-PDF of the resolved region. We adopt a direct $\chi^2$ fit
to this column density distribution as our reference $N$-PDF.
The aim of this test is to demonstrate the method's ability to retrieve a non-ideal
$N$-PDF, given a simple averaged emission function and noisy
observations. We also use it to explore the degeneracies inherent
to beam mixing.

The Bayesian inversion is carried out using the setup described in Section~\ref{sec:bayesian-inversion}. We assume a multiplicative error of 10\%, which accounts for calibration uncertainties of the 30m/EMIR and NOEMA instruments. The additive error level is estimated from the ORION-B data cubes. For each spectrum, we compute the RMS noise level ($\sigma_{RMS}$) as the standard deviation of 220 signal-free channels. Since the input observations of the Bayesian inversion are integrated intensities, the estimated noise is propagated through the intensity integration to produce a map of the additive error level on the integrated intensities. The spatial average of the noise for each line is used as the additive error level. Both the optimisation and sampling processes run for 10\,000 iterations, which is enough to converge to a solution and fully sample the posterior PDF. 

\subsection{Method performance on ORION~B}

Figure~\ref{fig:otrionb-post-pdf} compares the true column density
PDF, the $\chi^2$ fit that we use as a reference, the Bayesian MAP estimation and
the individual $N$-PDFs computed from the MCMC samples. The parameters of the reference and Bayesian MAP estimated $N$-PDFs are summarised in Table~\ref{tab:orionb_npdf}. The Bayesian MAP estimation from the 3\,mm averaged line observations is in excellent agreement with the reference $N$-PDF over more than two orders of magnitude in column density. \modifr{The main difference is the width $\sigma$ of the LN which appears slightly underestimated compared to the true $N$-PDF. As pixels below $N_{\rm{H_2}} = 10^{21}$\,cm$^{-2}$ do not contribute to the total emission, the estimated $N$-PDF is slightly narrower, with a pixel area filling factor of 0.90 corresponding to the proportion of pixels with $N_{\rm{H_2}} \ge 10^{21}$\,cm$^{-2}$ in the ORION-B data}. We quote the $16-84^{\rm th}$ percentiles of the
MCMC $N$-PDF realisations as the uncertainty on each of the parameters. The dispersion among the MCMC $N$-PDFs is typically larger at low (\NH{} $< 2\times10^{21}$\,cm$^{-2}$) and
high (\NH{} $> 10^{22}$\,cm$^{-2}$) column densities.

Figure \ref{fig:orionb-post-pdf-corr} shows the
posterior distribution of the model parameters.  The mean column density
($N_0$) of the LN part of the distribution is the best constrained
parameter with all sampled $N$-PDFs peaking around $\sim
2.5\times10^{21}$\,cm$^{-2}$. At densities above $N_0$, all the MCMC
$N$-PDFs are consistent with the reference $N$-PDF, although some sampled
$N$-PDFs reproduce the PL part with a very large LN distribution.

More quantitatively, the Bayesian estimation $N_0 = (2.49^{+0.26}_{-0.22})
\times 10^{21}$ closely matches the reference value of $N_0 = 2.38 \times
10^{21}$ cm$^{-2}$. The PL index $\alpha$ is likewise tightly constrained
and accurate, with an estimated value of $\alpha$ $=3.04^{+0.25}_{-0.90}$,
compared to the reference value of $\alpha = 3.19$. The width $(\sigma)$ of
the LN part of the distribution is more uncertain, $\sigma$
\modifr{$=0.40^{+0.39}_{-0.15}$}, \modifr{and slightly underestimate} the reference value
of $\sigma$=0.47. The column density of the transition from the LN to PL
part is also uncertain, with $N_\mathrm{thresh} =(5.23^{+40.53}_{-1.99})
\times 10^{21}$ cm$^{-2}$, with a strong tail in the posterior PDF. The MAP
estimation nonetheless closely matches the reference value of
$N_\mathrm{thres}$$ = 4.08 \times 10^{21}$ cm$^{-2}$.
\modifr{The pixel area filling factor is unbiased, with $\eta$ $=0.91^{+0.04}_{-0.28}$ compared to the reference value $\eta = 0.90$, corresponding to the proportion of pixels with \NH{} $> 10^{21}$\,cm$^{-2}$ in ORION-B. The pixel area filling factor being a scaling parameter, its uncertainty is directly related to the multiplicative error standard deviation of 10\%, which explains the wide posterior distribution of this parameter.}
\modifrr{In this high signal-to-noise ratio example, there is little to no degeneracy between $N_0$ and $\eta$.}


\subsection{Method limitations}

The wider posterior PDFs of $\alpha$, $\sigma$ and $r_{\rm{thresh}}$ = $N_{\rm{thresh}}$/$N_0$ illustrate a loss of information inherent to beam-averaging, which limits the precision of the parameter estimation.
Indeed, averaging the $N$-PDF produces two main effects on the model prediction and parameters.
First, the method is subject to degeneracies, i.e., a co-variation of several parameters can lead to a similar $N$-PDF shape, and therefore yield similar emission line predictions.
Second, the predicted intensities may lack sensitivity to a given parameter, due to the emission function that modulates the $N$-PDF that is then averaged to predict intensities. 
This occurs, for example, when a parameter primarily affects a part of the $N$-PDF where the emission function is negligible, such that the predicted intensities become insensitive to variations of this parameter. Alternatively, a large parameter variation might produce negligible variations in the $N$-PDF shape, and thus will not affect the predicted intensities. In this case, the fit procedure is also insensitive to the parameter.
In Appendices~\ref{sec:appendix:parameter-study} and~\ref{sec:appendix:degeneracies}, we present a detailed study of these effects. In the rest of this subsection, we summarize the main caveats that we established from our inversion of the ORION-B dataset.

The most important caveat is that a co-variation of several parameters can yield an overall similar $N$-PDF shape. This scenario is visible in the joint
posterior PDF of $\sigma$ and \modifr{$r_{\rm{thresh}}$}, which displays a clear
positive correlation. In this case, the 
power-law tail is comparable to the LN distribution before the transition point
$N_{\rm{thresh}}$. Hence, a larger LN distribution with a larger
$N_{\rm{thresh}}$ resembles a smaller LN distribution with a smaller
$N_{\rm{thresh}}$. Overall, the larger the LN distribution, the further
away the transition to PL tail must be to preserve a similar shape. This
results in the positive correlation between $\sigma$ and $r_{\rm{thres}}$
visible in Fig.~\ref{fig:orionb-post-pdf-corr}.
Figure~\ref{fig:degeneracies:sigma-rthres} illustrates this
effect.

We further observe that compensation effects between the shapes of the $N$-PDF
and the line emission function cause a "T" shape of the joint
posterior PDF of $\alpha$ and $\sigma$. For $\sigma$ values larger than the
reference value, $r_{\rm{thresh}}$ increases and so does the uncertainty on $\alpha$, which causes the horizontal top spread of the joint posterior
PDF. Indeed, as $N_{\rm{thres}}$ increases, the PL tail of the $N$-PDF becomes
smaller. This in turn decreases the fraction of the beam-averaged emission
sensitive to the PL tail to the point where the predicted intensities are no longer sensitive to the $\alpha$ parameter. This is illustrated in
Fig.~\ref{fig:degeneracies:alpha}
and~\ref{fig:emission-function:rthres-alpha}.

For $\sigma$ values lower than the reference, the PL tail dominates and the
LN width $\sigma$ becomes unconstrained, which produces the vertical
spread. Orion~B's $N$-PDF exhibits a strong PL component, which is visible
even at low $N_\mathrm{thres} = 4.08 \times 10^{21}$ cm$^{-2}$ relative to
$N_0 = 2.38 \times 10^{21}$ cm$^{-2}$. As a result, variations of $\sigma$
mostly affect column densities below $\sim 1 \times 10^{21}$ cm$^{-2}$,
where emission from all lines is below the noise level. Consequently,
$\sigma$ is poorly constrained and its posterior PDF is dominated by the
uniform prior. This effect is illustrated in Fig.~\ref{fig:degeneracies:sigma}. 
There is, however, a clear break in the posterior PDF at $\sigma
= 1$ even though the validity interval goes up to $\sigma = 2$. We
attribute this to our transition slope criterion: a $\sigma$ value greater
than one would result in a tangent at the transition flatter than the PL
index $\alpha$ of Orion~B, violating the slope criterion that we impose.

\section{Application to M51 observations}
\label{sec:results-M51}



The Surveying the Whirlpool at Arcseconds with NOEMA (SWAN) survey
\citep{Stuber2025} is a 3\,mm line survey of the M51 galaxy over a $5 \times
7$kpc FoV at \modifrr{140}\,pc resolution \modifrr{(pixel size of 45\,pc)}.  The dataset includes high
sensitivity observations of \nhp{}, as well as \thico{}, \eigco{}, \hcn{},
\hnc{} and \hcop{}. 
Here, we analyse a $700
\times 700$\,pc region within the SWAN FoV centred on M51's inner western
spiral arm.
The spiral arm crosses our test region from the north-east to the south-west (see Figure~\ref{fig:observations_predictions}).  The signal from many lines is faint outside the arm, except for a patch of brighter
emission in the north-west corner. We use the same setup for the Bayesian inversion procedure as for our test on the ORION-B data. 

\subsection{Spatial distribution of the $N$-PDF parameters}
\label{sec:results-M51:parameters}

\ParamsMap{}


\modifj{Figure~\ref{fig:m51:param-map} displays the spatial distribution of the
estimated parameters: the mean of the LN part $(N_0)$, \modifr{the pixel area filling factor ($\eta$), the average density including blank sky contributions,} the width
$(\sigma)$, the location of the LN to PL
transition $(N_{\rm thresh})$ and the PL index $(\alpha)$. White pixels represent lines of sight where the
integrated intensity is negative for at least one of the observed
lines. Overall, there is clear spatial structure in the $N$-PDF parameter maps of our test region,
with a transition from strong, high density PL tails in the spiral arm to
lower density, \modifr{mostly} LN $N$-PDFs outside the arm. Some of the pixels adjacent to the white pixels show a
pronounced PL tail, but the S/N is low at these
positions and the fitted parameters are quite uncertain.}


\modifr{More quantitatively, the LN mean column density appears relatively constant accross the FoV at $N_0\sim10^{22}$\,cm$^{-2}$, with the exception of lower $N_0$ values west of the spiral arm and higher values in low S/N regions. \modifrr{These higher $N_0$ values coupled with very low pixel area filling factors are due to a level of degeneracy between $N_0$ and $\eta$ arising in low S/N pixels, as described in Appendix~\ref{sec:appendix:degeneracies}}. The pixel area filling factor on the other hand displays sharp contrasts, going from unity in the spiral arm to $\sim$0.1 outside. The average column density including blank contributions (i.e., $\langle \Sigma \rangle \times \eta$) is consistent with the observed \tweco{} emission. This average column density is enhanced in the spiral arm, reaching up to a few  $10^{22}$cm$^{-2}$, and rather uniform outside, with $N_0\sim5\times10^{21}$cm$^{-2}$ on average and down to $10^{21}$cm$^{-2}$.
The $N_{\rm thresh}$ distribution is bimodal, with ($N_{\rm thresh} \sim2\times N_0$) in the spiral arm, and $N_{\rm
thresh} \geq 100\,N_0$ outside. This reflects the presence and decay of the
PL tail, inside and outside the spiral arm, respectively.  
The PL index is rather constant and
within the $[2,4]$ interval expected for gravitational collapse inside the
spiral arm. 
We observe a slight flattening of the PL from inside to outside of the spiral arm. 
We consider the $\alpha$ values estimated outside the spiral arm region to be unreliable since the $N$-PDFs are mostly LN.
The LN width $\sigma$ is moderate in the spiral arm, broadly consistent with Milky Way cloud values \citep[$\sigma\sim0.5$, e.g.,][]{Schneider2022}. Except for low S/N pixels, the highest $\sigma$ values are associated with low $N_0$ values, located in the north-western part of the FoV. This combination of high $\sigma$ and low $N_0$ effectively produces a low emission area filling factor, that is most of the $N$-PDF is below the threshold for \tweco{} emission, hence consistent with atomic or diffuse, CO-dark partially molecular gas.
, even though the pixel area filling factor is close to unity. The north-eastern part of the FoV displays the lowest $N$-PDF widths, with $\sigma\sim0.1$, hinting that turbulence might be lowly supersonic or lowly compressive.}


\subsection{Spatial distribution of dense gas}
\label{sec:results-M51:dense:gas}
\MdenseMap{}

\modifj{Figure~\ref{fig:m51:Mdense-map} compares the spatial distribution of the
dense gas mass, PL gas mass, and 24\,$\mu$m emission, which we use as a proxy for
star formation. The dense and PL gas mass are computed from the estimated
$N$-PDF parameters following the equations presented in
Section~\ref{sec:model:parameters}, and adopting a distance to M51 of $d =
8.58$\,Mpc.}

\modifj{The spatial distributions of dense and PL gas mass are similar. The mass
is up to 10 times larger inside the spiral arm compared to regions outside the arm. A concentrated region of higher masses is visible in the centre of the
arm, while moderately high dense gas masses are present in the south-western
part of the spiral arm. For pixels outside the arm, there is little to no mass in the dense or PL regime.  The transition from LN to PL occurs around $\sim
2\times10^{22}$cm$^{-2}$, which is slightly higher than the critical density of
filaments ($N_{{\rm H_2}} = 7\times 10^{21}\,$cm$^{-2}$) determined by
\cite{Andre2014} and the dense gas threshold derived by \cite{Lada2010}
over which the SFR is observed to be proportional to dense gas mass in
local clouds. The PL gas mass map shows a slightly higher contrast than the
dense gas mass map. In the north-east
of our target region, where the emission is associated with LN
$N$-PDFs, the dense (and PL) gas mass fraction is low, despite the bright emission associated with this region.}

\modifj{The pixels that are coincident with the peak of 24$\mu$m emission in the spiral arm exhibit $N$-PDFs with a pronounced PL tail and a large dense gas mass fraction. We identify pixels with $N$-PDFs with $\alpha \in [2.5,5]$ (see Sec~\ref{sec:model:parametrisation:PL}), and $f_\text{grav} > 25\%$ as being susceptible to gravitational collapse. The first condition is a 
standard criterion (see Sec~\ref{sec:model:parametrisation:PL}), while the second condition ensures that the PL represents a large fraction of the total emission, and thus that the $\alpha$, $r_{\rm{thresh}}$ and, by extension, $M_{\rm{PL}}$ estimates are reliable. 
We present this relationship more directly in Fig.~\ref{fig:m51:mdense-sfr-corr}, where we show the correlation
between the 24\,$\mu$m surface brightness and the dense and PL gas mass. We consider
two subsets of the data, depending on whether a pixel exhibits an $N$-PDF consistent with gravitational collapse as defined above. Both the dense and the PL gas mass in the gravity-dominated pixels (red symbols) demonstrate a tight linear correlation with 24\,$\mu$\,m surface brightness, with a correlation coefficient $r \sim 0.8$, whereas pixels that do not fulfill our criteria for gravitational collapse (greyscale symbols) show no such correlation. The intensities of individual emission lines such as \hcn{} or \nhp{} display weaker and flatter ($r = 0.7$, $s = 0.6$) correlations with the 24$\mu$m surface brightness than the dense and PL gas mass.}




\MdenseSFR{}

\subsection {Interpretation}
\label{sec:discussion:interpretation-M51}

%

\modifj{Our M\,51 target region contains a spiral arm segment with bright
3\,mm line emission, with fainter, more diffuse
emission arising in the interarm region. The north-eastern corner of our field, i.e., the interarm pixels situated closest to the centre of M51, also exhibits moderately bright emission.}

\modifj{Our method delivers several interesting results. First, the spiral arm
hosts the highest dense gas masses, up to $M_\text{dense} = 10^6$M$_\odot$. Second, the $N$-PDFs located in the spiral arm show \modifr{average} LN distributions ($\sigma < 0.6$), suggestive of \modifr{supersonic} turbulent gas with average Mach numbers, combined with pronounced PL tails with indices ranging
from 2.5 to 4 consistent with gravitational collapse. 
}
\modifj{Third, pixels where a pronounced PL tail is present display a strong,
approximately linear correlation between the dense and PL gas masses and the 24$\,\mu$m emission, which we consider here as an embedded star formation tracer. Conversely, pixels devoid of PL tails show
no such correlation. A linear correlation between dense gas mass and a star
formation tracer is consistent with the hypothesis that the star formation
efficiency is constant above a given column density threshold as observed
in nearby Milky Way clouds \citep{Lada2010}.}


\modifj{Fourth, the transition from LN to PL inside the spiral arm occurs at a roughly constant column density,
$N_{\rm{thresh}} \sim 2\times10^{22}$\,cm$^{-2}$. This value is similar to the empirical dense gas threshold above which star formation is observed to be constant in nearby Milky Way clouds
\citep{Lada2010}. Fifth, the inferred PL indices appear correlated with 24$\mu$m emission, becoming steeper as the  24$\mu$m surface brightness increases. This result is not entirely expected: regions with higher star formation activity should be associated with flatter PLs since the SFR is proportional to the available dense gas mass \citep{Lada2010}. In
magnetohydrodynamical simulations without stellar feedback, \cite{Federrath2013} find that molecular clouds with flatter PL
tails have a higher star formation efficiency (where $SFE = M_\star /
(M_{cloud} + M_\star)$. The trend that we observe could be
the result of feedback mechanisms that are efficient in disrupting high density gas. In embedded regions, these mechanisms could preferentially act on gas with column densities corresponding to the PL tail of
the PDF, steepening the PL.}

\modifj{Finally, outside the spiral arm, the $N$-PDFs are mostly LN, with \modifr{smaller}
widths \modifr{($\sigma<0.3$)} than in the spiral arm. We interpret this as
evidence that the molecular gas outside the arm is turbulence-dominated, \modifr{lowly-supersonic,}
and largely stable against gravitational collapse.}

\section{Discussion}
\label{sec:discussion}

\subsection {Comparison with the line ratio approach}
\label{sec:discussion:comparison-ratios}

To date, most studies of the dense molecular gas in external galaxies have relied on a line ratio approach.
This approach can be likened to a two-zone model, with a low-density gas component, typically $n_{\rm{H_2}}$ < 10$^{4.5}$ cm$^{-2}$ \citep{GaoSolomon2004}, and a dense gas component at higher densities. The main assumption of the line ratio approach is that the emission of certain lines arise predominantly from the dense component \citep[e.g., \hcn{}][]{GaoSolomon2004}. Such lines are labelled as ''dense gas tracers''. Emission lines such as \tweco{} that trace both low and high gas density components are regarded as ''bulk gas'' tracers. 
Following these assumptions, combining observations of a bulk gas tracer and a dense gas tracer provides information about the dense gas fraction. By further assuming that these tracers vary linearly with the average density of the gas that they trace (i.e., assuming a constant $\alpha_{\rm{CO}}$ or dense gas conversion factor), then the dense gas fraction simply becomes the ratio between the dense gas tracer and the bulk gas tracer. 
Attractive for its simplicity, this approach has been shown to be effective for the interpretation of extragalactic observations \citep[see e.g.,][Chapter 3, for a recent overview]{Schinnerer2024}, where detecting a diverse suite of density-dependent emission lines is prohibitive with current observational facilities. 

\begin{figure*}
    \centering
    \includegraphics[width=0.8\linewidth]{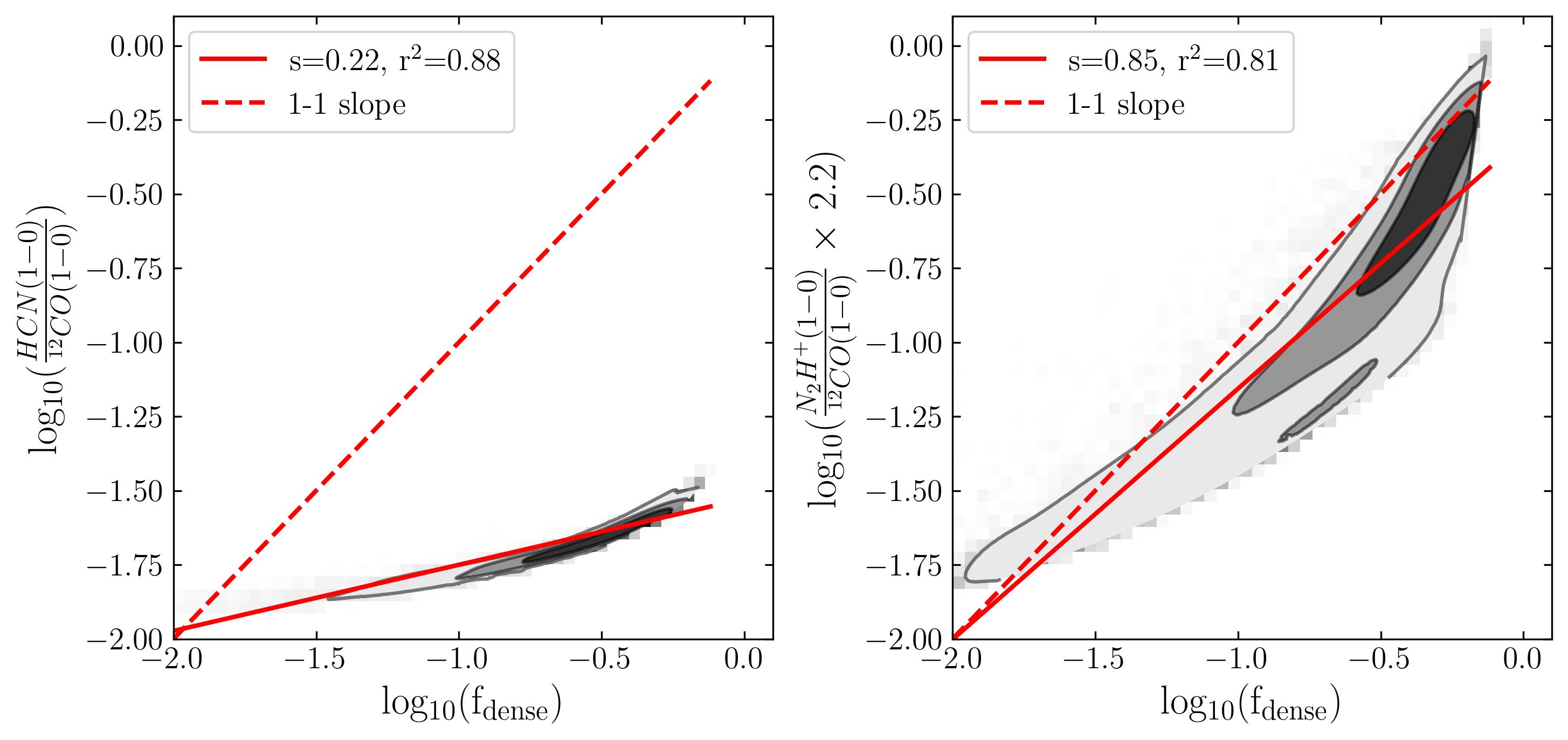}
    \caption{Beam-averaged line ratios between \hcn{} (\textit{left}) and \nhp{} (\textit{right}) over \tweco{} as a function of $f_{\rm{dense}}$, for a sample of 100 000 $N$-PDF with parameters uniformly sampled over $N_0 \in [5\times 10^{20} : 10^{22}]$, $\alpha \in [2.5 : 5]$, $\sigma \in [0.2 : 1.3]$ and $r_{\rm{thres}} \in [2 : 100]$. The dashed red line shows a linear relation, while the full red line shows a linear fit to the data. This parameter space is more limited than the validity intervals of Table~\ref{tab:val_intervals}, but more representative of $N$-PDFs observed in local clouds \citep{Schneider2022}.}
    \label{fig:fdense_ratios_model}
\end{figure*}

The method presented in this paper was designed to alleviate these shortcomings of the line ratio approach, leveraging the wide bandwidth, multi-line capacity of modern backends and recent observations revealing how the integrated line emission varies as a function of column density at high ($\sim 0.01$\,pc) resolution. 
For comparison, Figure~\ref{fig:fdense_ratios_model} displays the \hcn{} and \nhp{} ratios over \tweco{} as a function of $f_{\rm{dense}}$ for a sample of 100 000 $N$-PDFs uniformly covering the parameter space.
While the correlation between the \hcn{} over \tweco{} ratio and $f_{\rm{dense}}$ is strong (Pearson correlation coefficient of $r^2 = 0.88$), it is significantly sub-linear, with a correlation coefficient of $m=0.22$. This is qualitatively consistent with the results of \cite{Bemis2024}, and again demonstrates that the \hcn{} over \tweco{} ratio does not straightforwardly trace the fraction of gas above $N_{\rm{H_2}} = 10^{22}$\,cm$^{-2}$. The significant contribution of low ($N_{\rm{H_2}} < 10^{22}$\,cm$^{-2}$) density gas to the total \hcn{} emission attenuates and flattens the correlation between dense gas mass and \hcn{} emission \citep{Leroy2017, Bemis2024}.

On the other hand, the ratio between \nhp{} emission and $f_{\rm{dense}}$ is strong ($r^2=0.81$) and almost linear ($m=0.84$) over two orders of magnitude in line ratios and dense gas fraction.  Thus, the ratio between \nhp{} and \tweco{} appears to be a significantly better tracer of the dense gas fraction. The obvious practical drawback is that this diagnostic ratio requires robust detections of \nhp{}, which often implies prohibitive integration times for extragalactic sources.

\subsection {Limitations and future improvements of the model}
\label{sec:discussion:limitation-function}

While they already give promising results, the parametrizations of the
   column density PDF and the emission function presented here are not
   universal and can be improved. The parametrisation of the N-PDF
   as a combination of a LN and a PL function, while both observationally and
   theoretically justified, assumes that the gas is isothermal. In the
   case of other polytropic indices, the (volume) density PDF can deviate
   from a pure LN, with for instance the development of a PL tail at low 
   densities \citep{Federrath2015b}. Empirical studies likewise suggest that the
   $N$-PDF can display a bimodal profile, with a peak at high and low column densities
   \citep{Schneider2022}. The low column density peak however corresponds to atomic gas, thus below the
   column densities that we probed using molecular gas emission lines. In the extreme case where the turbulent forcing is \modifr{strongly compressive}, recent studies have shown that the volume density PDF differs from a LN distribution and instead resembles a Castaing-Hopkins PDF \citep{Hennebelle2024, Brucy2024}, especially if the Mach number is large ($>20$). The difference between the LN and Castaing-Hopkins shape is significant at high densities, with less dense gas in the Castaing-Hopkins PDF compared to the LN for a similar Mach number \citep{Hennebelle2024}. Investigating the characteristics of turbulence using other parametric density PDFs will be the subject of future work.

The emission function used for the analysis in this paper can also be improved. As an empirical law fitted on high quality data, our current emission function encapsulates the physics and chemistry relevant for molecular line emission across the range of column densities present in the ORION-B data. The drawback is the model's lack of universality, making it most suitable for interpreting unresolved observations of regions where the gas properties are similar to those in Orion~B. Variations in several potentially important physical parameters, such as the amplitude of turbulent gas motions and gas kinetic temperature, are not currently taken into account. 
The former could have a significant impact on the emission from optically thick lines such has \tweco{}. Temperature variations, on the other hand, would mostly affect line emission arising from the outer layers of the cloud, e.g., \tweco{} or \thico{}. Emission lines such as \nhp{} that are excited in the colder interior of the cloud should be less subject to temperature variations.


Beyond radiative transfer, chemical effects are also expected to influence the line emission, and could be incorporated into our emission model to some extent. For example, the photodissociation front of the cloud is expected to vary depending on the gas-phase metallicity and radiation field illuminating the molecular gas \citep{Kaufman1999}. A deeper front would globally shift the emission function displayed in Figure~\ref{fig:emission-function} to the right, progressively decreasing the emission from lower (\tweco{}) to higher (\nhp{}) density lines. 
Similarly, a higher temperature or stronger radiation field pervading deeper into the cloud could shift the freeze out of molecules onto dust grains to higher column densities. A higher or lower column or volume density threshold for freeze-out of carbonated molecules would then significantly alter the emission function of \nhp{}. 
Furthermore, FUV radiation can enhance the formation of HCN as well as increase the ionization fraction \citep{Santa-Maria2023}. The HCN molecules can then be significantly excited by e$^-$ collisions. These effects can produce low-level, extended  \hcn{} emission in diffuse, illuminated gas. Averaged over an entire cloud, this emission arising from lower density (n$_{\text{H$_2$}}<10^{4}$\,cm$^{-3}$) gas can account for more than half of the total \hcn{} \citep{Santa-Maria2023}.
Cosmic rays and their ionization rates are another key parameter left out of the equation for now. They play a major role in the intricate chemical network regulating the abundances of complex molecules. For instance, regions exposed to elevated ionizing rates such as galactic nuclei can display extended and overabundant \nhp{} emission \citep{Santa-Maria2021} compared to Orion~B.
Finally, stellar nucleosynthesis of carbon and oxygen isotope can change the relative abundances of CO isotopologues. $^{12}$C and $^{16}$O isotopes are preferentially produced in massive, short lived stars, whereas $^{13}$C and $^{18}$O are produced in lower-mass, longer lived stars during a slower process. As the population of stars varies from mostly young, massive stars in  the inner star-forming disk to older, lower mass stars in the outer disk, both \thico{} over \tweco{} and \eigco{} over \tweco{} are expected to increase as a function of galactocentric radius \citep{Wilson1994}.

The logical next step to develop our method would be to implement new physical parameters in the emission function. This would make the model more flexible while retaining its empirical motivation and permitting benchmarks against high resolution observations. 
\modifj{There are several avenues that could be explored to implement new parameters: 1) discretizing
  the emission model into several one-zone models for specific parameters
  (e.g., an inner and outer cloud temperature, \citep[see e.g.,][]{Segal2024}, 2) defining profiles of relevant parameters as a function of the
  column (or volume) density \citep[e.g.,][]{Tafalla2021}, or 3) a combination of the two
  previous approaches.}
A straightforward extension of the current model would be to impose a specific temperature corresponding to the cloud's surface for optically thick tracers (e.g., \tweco{}), and a temperature profile as a function of column density for emission lines that are excited in the colder and denser parts of the cloud. 

An alternative way to improve the emission function would be to express the average intensity as a function of a two-dimensional PDF of column density and temperature (or other relevant parameter). However this would also require the two-dimensional PDF to be parametric and quickly increase the number of parameters to be estimated. For instance, a temperature PDF composed of a combination of PLs would introduce three additional parameters to be estimated by the inversion. 

In all of the above cases, the improvement of the emission function requires adding parameters. The main obstacle towards implementing this increase of physical parameters remains the limited number of emission lines that are accessible to observations of typical extragalactic star-forming regions. The number of independent observations must be larger than the number of parameters for the model to be invertible and to alleviate degeneracies. Additionally, the observed emission lines should be sensitive to the different physical parameters. To include more radiative transfer parameters, such as temperature, opacity or even abundance variations, the natural next step would be to add higher order J-lines. The theoretical maximum level of information that an observable can deliver about a physical parameter can be quantitatively estimated using the mutual information measurement \citep[][sect. 8.6]{cover2006elements}. This metric is particularly well-suited to informing the selection of lines depending on the physical parameters of interest, as shown by \cite{Einig2024}, and we encourage its wider use for observation planning.

\section{Summary and conclusions}
\label{sec:conclusion}

This paper presents a new method to infer the column density PDF and gas
mass above a column density threshold from a set of spatially unresolved 3\,mm
emission line observations.  The method assumes that the column density PDF
can be parametrised as a combination of a lognormal distribution due to
supersonic isothermal turbulence, and a power law distribution at higher
densities, the latter arising from gravitationally unstable or collapsing
structures. It also assumes that the intensity of the observed 3\,mm
emission lines depends mostly on the column density. The emissivity of each line is
calibrated on the ORION-B data.

For validation, the method was applied to the spatially and spectrally averaged
ORION-B data for nine emission lines. In this case, the method accurately
recovers the column density PDF of the resolved data, and the model
reproduces the integrated intensities of the input emission lines within
noise and calibration uncertainties. We then applied the method to
3mm line observations towards a $700\times700$\,pc field covering a
spiral arm segment in the M51 galaxy from the SWAN Large Programme. The
main results of our analysis of this region are the following :
\begin{enumerate}
\item The model reproduces the observed integrated intensities within
  \modifr{25}\% on average. With our current model, the predicted integrated
  intensities of the \thico{} and \eigco{} lines for some pixels show the
  largest discrepancy (up to a factor of 2) with the SWAN observations.
\item $N$-PDF power laws, consistent with gravitational collapse, are
  detected almost exclusively in the spiral arm region that crosses our M51
 field. Outside the spiral arm, the $N$-PDFs are predominantly lognormal,
  with a \modifr{small} width, consistent with \modifr{lowly-}supersonic \modifr{or lowly-compressive} turbulence
  dominating the gas dynamics.
\item The relationship between the inferred dense gas mass and the
  24$\,\mu$m emission varies as a function of the $N$-PDF properties. There is no correlation between these quantities for purely LN $N$-PDFs, while a strong, almost linear correlation is present for $N$-PDFs with a significant PL component. The correlation steepens from null to linear as the fraction of gas mass in the PL increases from $\le5\%$ to $\ge25$\% .
  %
\item The correlation between the 24$\,\mu$m emission and inferred gas mass in the power-law tail of the $N$-PDF is closer to linear and tighter than the correlation with the integrated intensities of individual emission lines. When the fraction of gas in the
  power-law tail is greater than 25\%, we find a correlation coefficient $r=0.85$ and slope $s=1$. Correlations with the integrated intensities of both \hcn{} and \nhp{} display weaker correlation coefficients ($r\sim0.7$) and flatter slopes ($s\sim0.6$).
\end{enumerate}


Efforts to expand the current study are already underway. 
We are working to improve the emission function to include additional 
physical effects such as variations in the kinetic temperature and linewidth by developing the modelling strategies presented in Section~\ref{sec:discussion:limitation-function}.  
The application of this improved model to the full SWAN dataset for M51 will be presented in a future paper (Zakardjian et al, in prep).

\begin{acknowledgements}
  This work is based on observations carried out under project numbers
  019-13, 022-14, 145-14, 122-15, 018-16, and finally the large program
  number 124-16 with the IRAM 30m telescope and large program number M19AA
  with the NOEMA interferometer. IRAM is supported by INSU/CNRS (France),
  MPG (Germany) and IGN (Spain). This research has also made use of data
  from the \textit{Herschel} Gould Belt Survey (HGBS) project
  (\url{http://gouldbelt-herschel.cea.fr}).
  This work received support from the French Agence Nationale de la
  Recherche through the DAOISM grant ANR-21-CE31-0010, and from the
  Programme National ``Physique et Chimie du Milieu Interstellaire'' (PCMI)
  of CNRS/INSU with INC/INP, co-funded by CEA and CNES.
  M.G.S.M. and J.R.G. thank the Spanish MICINN for funding support under
  grant PID2019-106110GB-I00.  M.G.S.M acknowledges support from the NSF
  under grant CAREER 2142300.
  Part of the research was carried out at the Jet Propulsion Laboratory,
  California Institute of Technology, under a contract with the National
  Aeronautics and Space Administration (80NM0018D0004).
  D.C.L. acknowledges financial support from the National Aeronautics and
  Space Administration (NASA) Astrophysics Data Analysis Program (ADAP).
\end{acknowledgements}
\bibliographystyle{aa} %
\bibliography{biblio.bib}

\begin{appendix}

\twocolumn
\section {Example model degeneracies}
\label{sec:appendix:degeneracies}

This section provides examples of common ''beam-averaging'' degeneracies, illustrating how $N$-PDFs with distinct parameters can yield similar average integrated intensities.

\begin{figure}[h]
    \centering 
    \includegraphics[width = 0.9\linewidth]{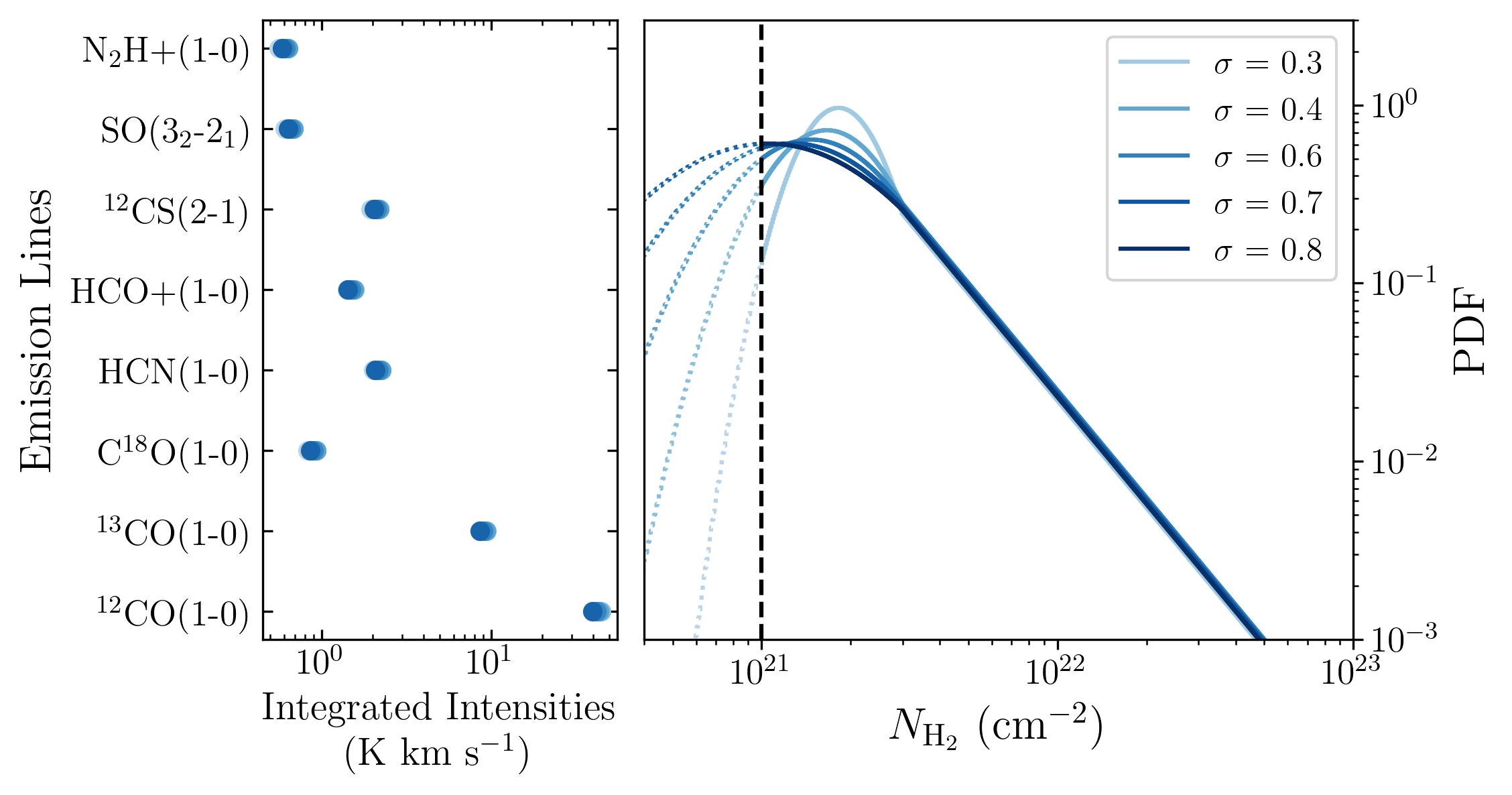}
\caption{\textit{Left:} Model predicted integrated intensities for a piecewise LN and PL $N$-PDF, with $N_0 = 2\times 10^{21}$\,cm$^{-2}$, $r_{\rm{thres}}$=1.5, $\alpha=3$ and $\sigma$ increasing from 0.3 to 0.8. All predicted integrated intensities vary by 10\% at most, which illustrates the degeneracy on the parameter $\sigma$ in the particular case of low $N_0$ and strong PL component. The right panels displays the corresponding $N$-PDF with varying $\sigma$. The dashed vertical line shows the column density limit below which the emission of all lines is less than 0.1 K\,km\,s$^{-1}$.}
\label{fig:degeneracies:sigma}
\end{figure}

Figure~\ref{fig:degeneracies:sigma} depicts a degeneracy on $\sigma$, which arises if the $N$-PDF is predominantly PL-like and centred at relatively low column densities ($N_{\rm{H_2}} < 5\times10^{21}$\,cm$^{-2}$. In this case, similar to Orion~B in Section~\ref{sec:results-orionb}, changing the LN width only affects the $N$-PDF where the emission function is almost null. The average integrated intensities are thus unaffected, which makes the $\sigma$ parameter degenerate.

\begin{figure}[h]
    \centering 
    \includegraphics[width = 0.9\linewidth]{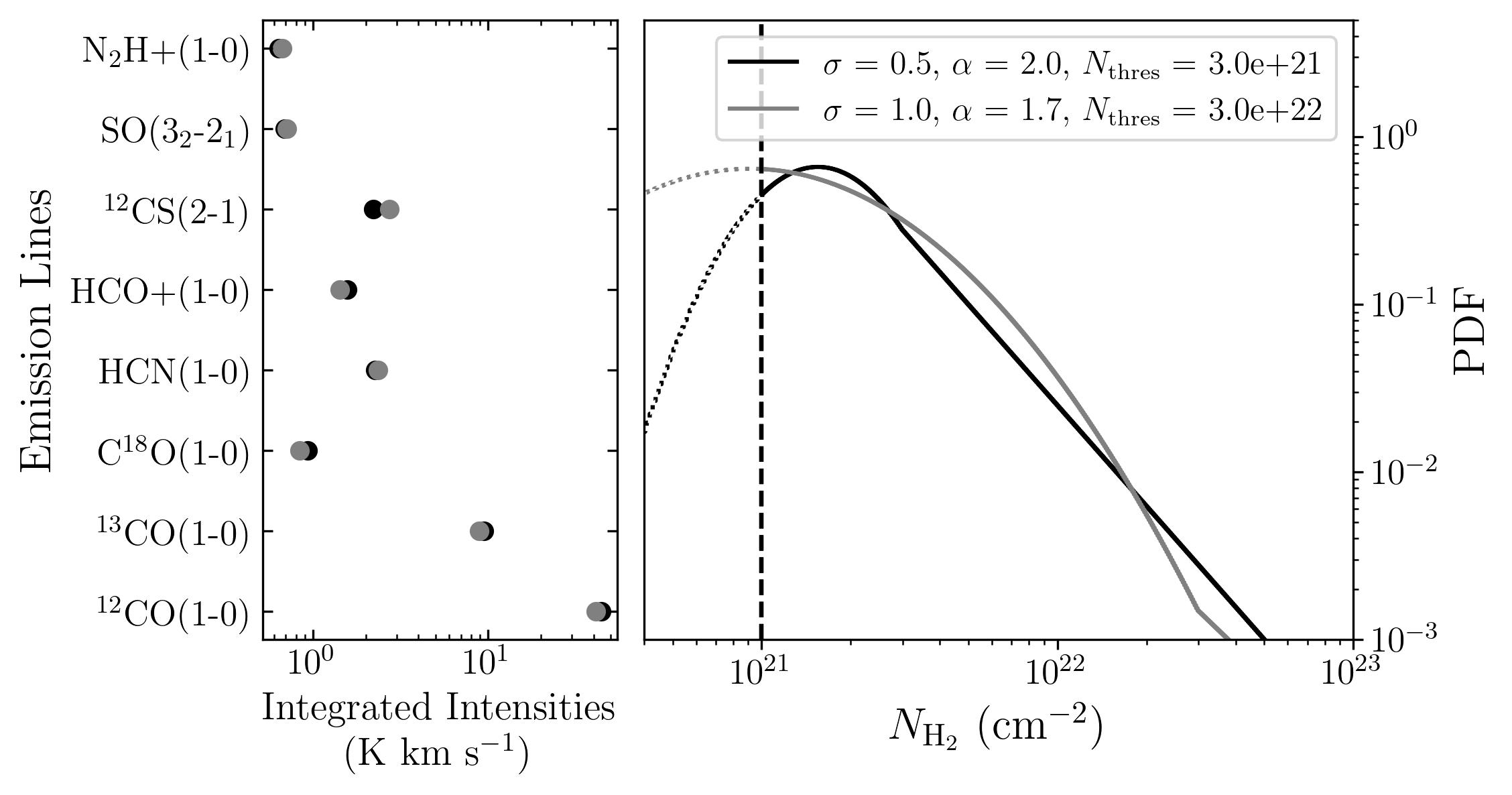}
\caption{Same as Figure~\ref{fig:degeneracies:sigma}, except two specific $N$-PDF are illustrated: the first has a narrow LN width but a strong PL, while the other has a large LN width but a smaller PL. This illustrates how two $N$-PDF with notably different parameters can appear alike and give comparable average integrated intensities.}
\label{fig:degeneracies:sigma-rthres}
\end{figure}

Figure~\ref{fig:degeneracies:sigma-rthres} illustrates a degeneracy between $\sigma$ and $r_{\rm{thres}}$, in the case where a co-variation of these parameters leads to a globally similar $N$-PDF shape, and therefore similar average integrated intensities. This kind of degeneracy appears as a correlation between $\sigma$ and $r_{\rm{thres}}$ in the posterior PDF of the Orion~B Bayesian inversion in Section~\ref{sec:results-orionb}.

\begin{figure}[h]
    \centering 
    \includegraphics[width = 0.9\linewidth]{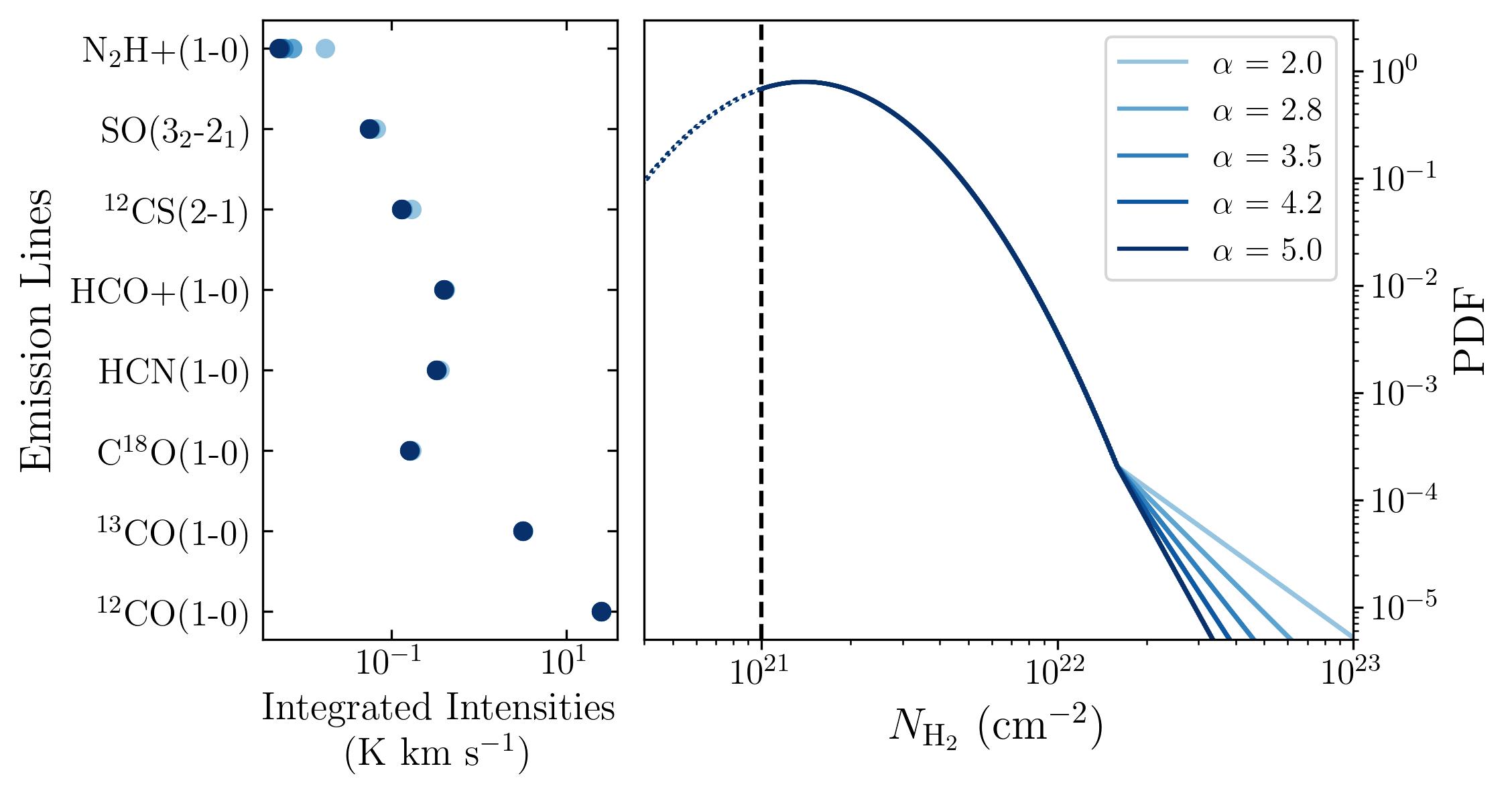}
\caption{Same as Figure~\ref{fig:degeneracies:sigma}, except the LN width is fixed to $\sigma = 1$ and the PL index is varying from 2 to 5. This illustrates the degeneracy on the parameter $\alpha$ when the PL appears far in the LN and therefore accounts for a negligible portion of the total $N$-PDF.}
\label{fig:degeneracies:alpha}
\end{figure}

Figure~\ref{fig:degeneracies:alpha} displays a typical degeneracy on the PL parameters in the case where the PL accounts for a small fraction of the $N$-PDF, even though it covers high densities and so high values of the emission function. Here, varying $\alpha$ does not change the average line intensities. In other words, the PL parameters become intractable if $r_{\rm{thres}}$ is high enough.

\begin{figure}[h]
    \centering 
    \includegraphics[width = 0.9\linewidth]{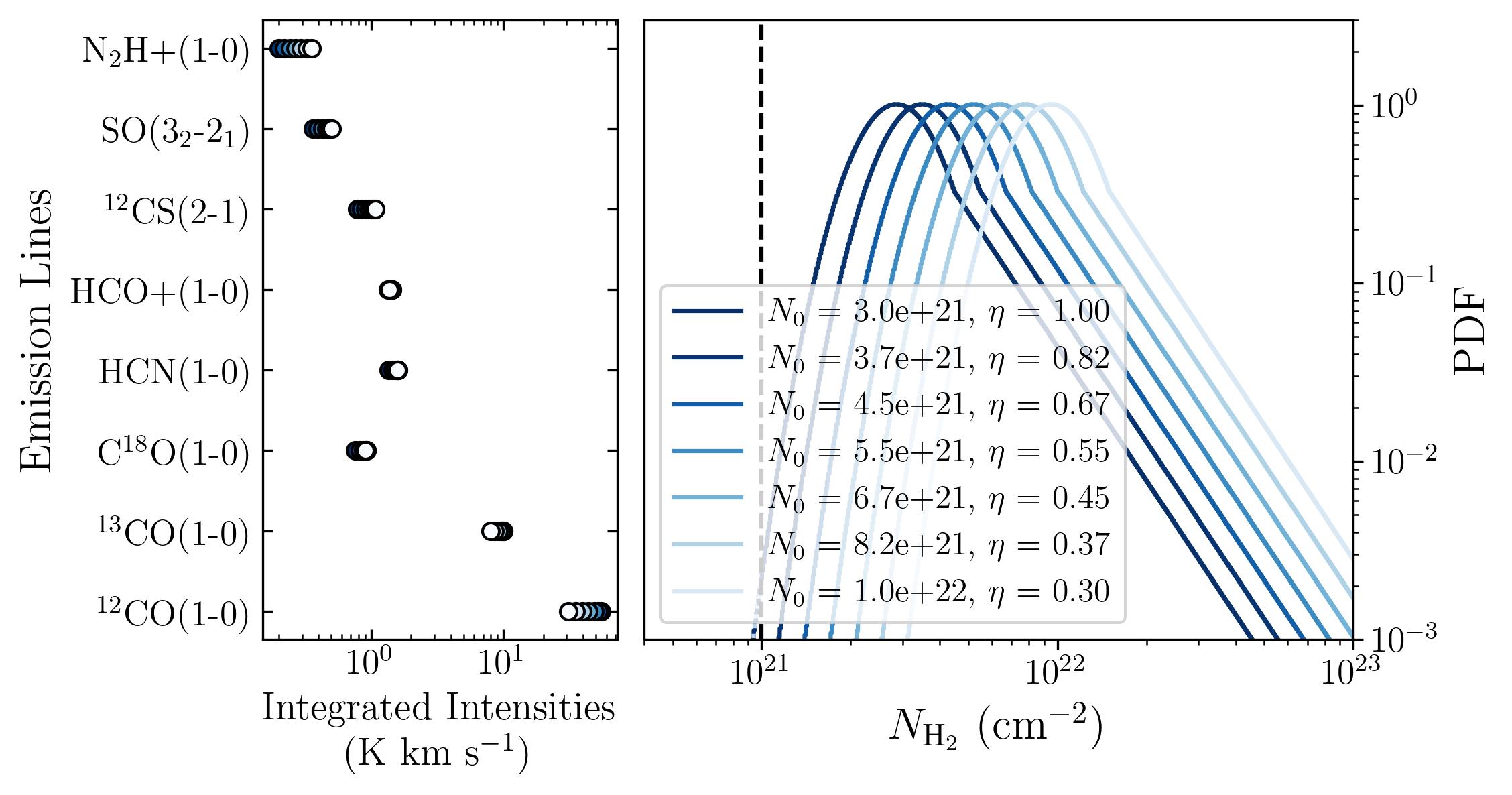}
\caption{\modifrr{\textit{Left:} Same as Figure~\ref{fig:degeneracies:sigma}, except for $N_0$ increasing from $3\times 10^{21}$\,cm$^{-2}$ to $1\times 10^{22}$\,cm$^{-2}$, $\eta$ decreasing from 1 to 0.3 and $\sigma = 0.3$. The predicted integrated intensities vary by 10\% to 50\% at most (30\% on average), which illustrates the degeneracy between the LN mean $N_0$ and the pixel area filling factor $\eta$. The right panels displays the corresponding $N$-PDF with varying $N_0$.}}
\label{fig:degeneracies:N0s_phi}
\end{figure}

\modifrr{Finally, figure~\ref{fig:degeneracies:N0s_phi} illustrates the potential degeneracy between $N_0$ and $\eta$. This figure shows 1) a set of $N$-PDFs with $N_0$ and $\eta$ are anti-varying over more than a factor three, and all other parameters held constant, and 2) the corresponding predicted line integrated intensities.
Overall, line integrated intensities vary by 5\% at least (\hcop{}) to 50\% at most (\nhp{} and \tweco{}).
While a predicted integrated intensity variation of 5\% is insufficient to lift the degeneracy between $\eta$ and $N_0$, a variation of 50\% will break this degeneracy as this intensity variation is much larger than calibration uncertainties.
However, this implies to well detect (at least) two lines sensitive to $N_0$ and $\eta$, such as \nhp{} and \tweco{}, which is consistent with peculiar $N_0$ values towards the lowest S/N pixels in the FoV.}

\section {Observations versus model predictions in Orion~B}
\label{sec:appendix:Orionb}

\begin{figure}[h]
    \centering
    \includegraphics[width=0.7\linewidth]{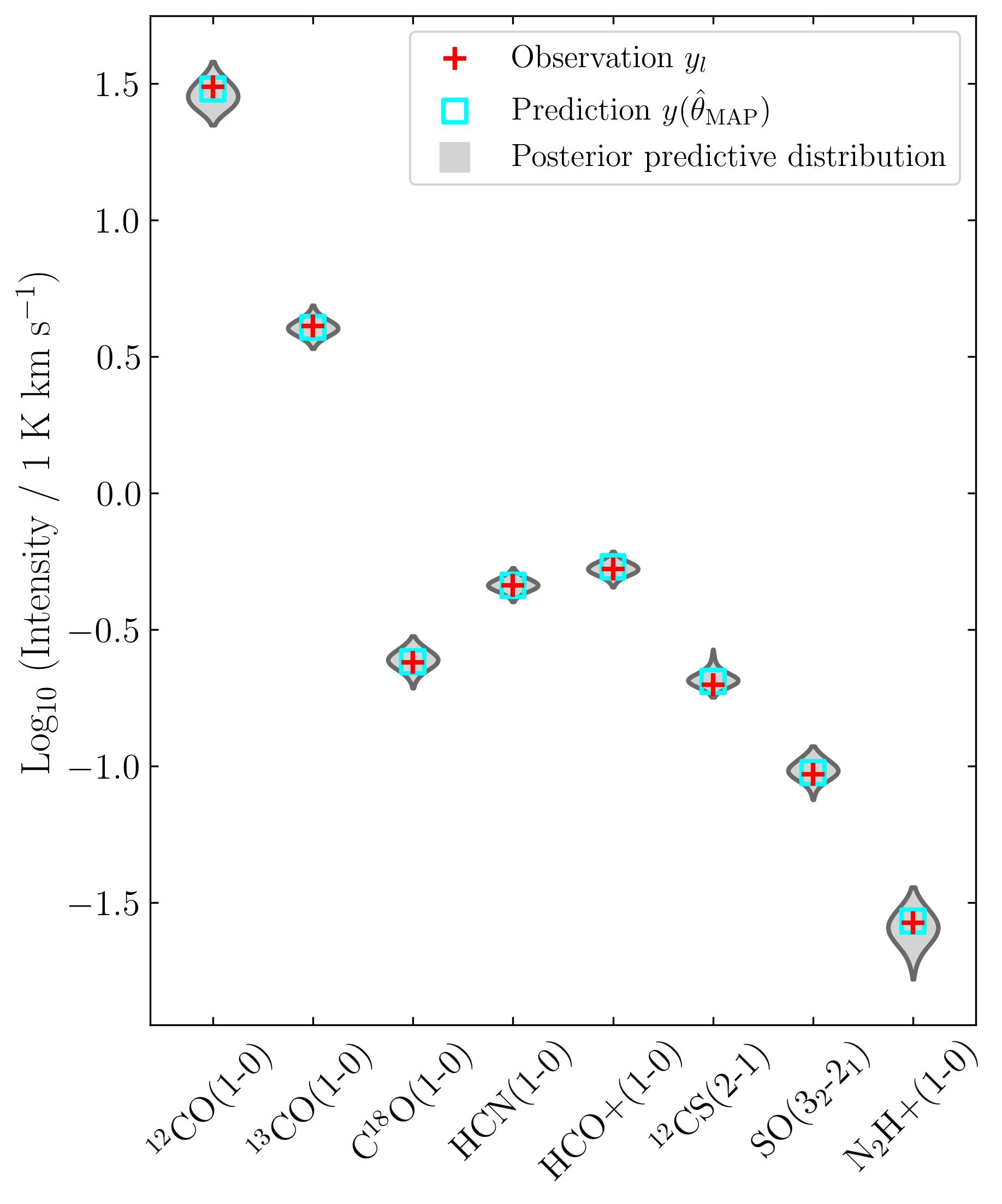}
    \caption{Posterior predictive distribution from the Bayesian inversion on the Orion~B data performed in Section~\ref{sec:results-orionb}. The red crosses show the observed intensity, the cyan squares show the model prediction for the MAP estimator, and the grey violin plot shows the histogram of model predictions for the entire posterior PDF (every MCMC iteration).}
    \label{fig:postpred_Orion-B}
\end{figure}

Figure~\ref{fig:postpred_Orion-B} shows the posterior predictive distribution from the Bayesian inversion on the Orion~B data performed in Section~\ref{sec:results-orionb}. The model prediction for the estimated parameters (MAP) matches the observed line intensities, and both the model best prediction and observations sit well within the distribution of prediction from the entire posterior PDF, with no visible offset.  

\begin{table}[h]
    \centering
    \caption{Orion-B $N$-PDF parameters estimated from $\chi^2$ fitting of the dust-derived column density map compared to the Bayesian inversion (MAP estimation) results.}
    \begin{tabular}{llllll}
    \hline
         Parameter & $N_0$ ($10^{21}$ cm$^{-2}$) & $\sigma$ & $r_{\rm{thres}}$ & $\alpha$ & $\eta$ \\
         \hline
         $\chi^2$ fitting & 2.38 & 0.47 & 2.43 & 3.19 & 0.90 \\
         MAP & $2.49^{+0.26}_{-0.22}$ & $0.40^{+0.39}_{-0.15}$ & $2.95^{+48.59}_{-1.29}$ & $3.04^{+0.25}_{-0.90}$  &
         $0.91^{+0.04}_{-0.28}$ \\
         \hline
    \end{tabular}
    \label{tab:orionb_npdf}
\end{table}

\onecolumn
\section {Observations versus model predictions in M51}
\label{sec:appendix:M51_obs}

\begin{figure*}[h]
    \centering
    \includegraphics[width=0.9\linewidth]{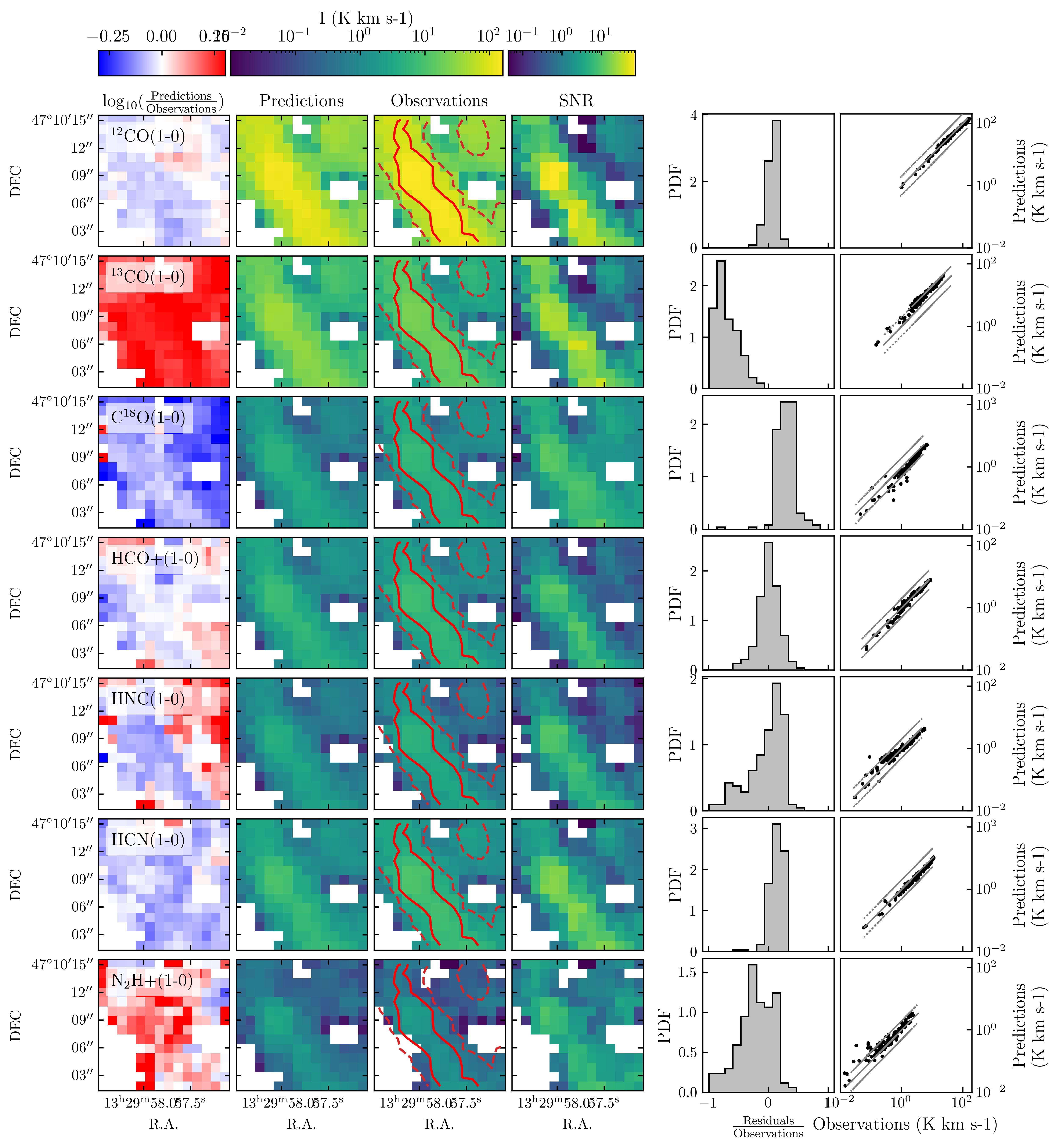}
    \caption{From the leftmost to the rightmost column: ratio of best model prediction over observed integrated intensities (in log$_{10}$ scale, the colour scale limits are a factor 2 in red and a factor 0.5 in blue),  best model predictions, observations, S/N level, histograms of the residuals normalised by the observations, and finally best model prediction as a function of observations. For the observation column, red contours in each panel indicate \thico{} integrated intensities of 4~and 12~K\,km\,s$^{-1}$ (dashed and solid contours, respectively). Each row is a different emission line, from top to bottom: \tweco{}, \thico{}, \eigco{}, \hcop{},  \hnc{}, \hcn{}, and \nhp{}.}
    \label{fig:observations_predictions}
\end{figure*}

\modifj{Figure~\ref{fig:observations_predictions} compares the model
     predictions to the observations across our M\,51 test region. The
     \tweco{} and \hcop{} integrated intensities are well-reproduced by the
     model, with an intensity weighted average prediction over observation
     ratio of 0.95 and 0.96, respectively. On one hand, the \hcn{} and
     \hnc{} integrated intensities are slightly underestimated, with
     average ratios of respectively 0.86 and 0.89. And the \eigco{}
     emission is noticeably underestimated by the model, with an average
     ratio of 0.72. On the other hand, the \thico{} \nhp{} intensities are
     consistently over-estimated by respectively $\sim70$\% and $\sim50\%$ on average.}

     \modifj{For most emission lines, a clear spatial pattern is visible in the
     residuals, with over-estimation of the line intensities in the arm and
     under-estimation outside, or vice versa.  Such a clear spatial pattern
     suggests local variations in the emission function, potentially caused
     by temperature, isotopologue abundance or opacity variations, as
     discussed in Section~\ref{sec:discussion:limitation-function}.}

\clearpage

\onecolumn
\section {Parametric study of the model}
\label{sec:appendix:parameter-study}

This section describes the model predictions and its potential degeneracies in terms of line ratio variations over two-dimensional projections of the parameter space. Here and in the following section, the term line ratios refers to the predicted integrated intensity of different lines relative to the integrated intensity of \tweco{}.

\subsection{LN: line ratio variations with $N_0$ \& $\sigma$}


Figure~\ref{fig:emission-function:N0-sig0} displays the line ratio variations as a function of $N_0$ and $\sigma$. For this grid of predictions, the distribution is purely LN, with $N_0$ varying from $10^{21}$ to $5\times10^{22}$\,cm$^{-2}$ while $\sigma$ increases from 0.2 to 1.8.

As expected, the parameter $N_0$ has the most impact on the observed line ratios. Its increase shifts the entire $N$-PDF to higher density values where the emission function predicts stronger intensities, while the overall $N$-PDF shape remains unchanged. As \tweco{} emission quickly saturates, all line ratios increase with $N_0$.
These line ratio variations with respect to $N_0$ range from one to four orders of magnitude, depending on the line considered. Emission lines for which the emission function flattens, or even saturates, at high $N_{\rm{H_2}}$ show less variation (e.g., \thico{}) than lines with steeper emission functions (e.g., \cs{} or \nhp{}).

Variations in the line ratios due to $\sigma$ are less pronounced, and mostly discernible for lower values of $N_0$. Indeed, in this case, increasing $\sigma$ will make the $N$-PDF sample progressively larger densities, increasing the line ratio values. However, if the $N_0$ value is high enough - that is, the entire $N$-PDF is centred on higher densities ($N{_{\rm{H_2}}}>10^{22}$\,cm$^{-2}$ - then increasing $\sigma$ induces less variation of the line ratios. This is because the emission functions of the different lines are significantly flatter at higher column densities. The increased degeneracy of $\sigma$ as $N_0$ increases is visible in Figure \ref{fig:emission-function:N0-sig0} as the ratio isocontours turning vertical, and is particularly pronounced for \hcn{}.

Overall, all the emission lines considered here vary strongly (by at least a factor four and up to three orders of magnitude) with either $\sigma$ or $N_0$ increasing by roughly an order of magnitude. In other words, for a pure LN distribution, both these parameters can be straightforwardly estimated.

\FigRatiosNSig{}

\subsection{Power law: line ratio variations with $R_{\rm{thresh}}$ \& $\alpha$}

Figure~\ref{fig:emission-function:rthres-alpha} shows the line ratios as a function of $r_{\rm{thres}}$ and $\alpha$ for a fixed LN distribution with $N_0 = 5\times10^{21}$\,cm$^{-2}$ and $\sigma$ = 0.6. $\alpha$ and $r_{\rm{thres}}$ vary from 1 to 3 and 2 to 15, respectively.

Qualitatively, all line ratios behave similarly. Increasing either $\alpha$ or $r_{\rm{thres}}$ brings the ratio values down. Increasing $\alpha$ steepens the PL part of the $N$-PDF, which reduces the range of higher densities covered and thus decreases the line ratios. Similarly, increasing $r_{\rm{thres}}$ will push the transition from LN to PL further into the LN distribution, reducing the contribution of the high density PL to the $N$-PDF. Reducing the high density tail of the $N$-PDF consequently decreases the line ratios.  

Quantitatively, there is overall little variation of the \thico{}, \eigco{}, \hcop{}, \hcn{} and \hnc{} ratios with $r_{\rm{thres}}$ and $\alpha$, with variations spanning less than an order of magnitude. Only \so{}n \cs{} and \nhp{} line ratios over \tweco{} vary by more than an order of magnitude over the parameter space. The \nhp{} over \tweco{} ratio in particular varies by more than two orders of magnitude and rather uniformly across the entire parameter space considered here. This makes \nhp{} the best tracer to estimate PL parameters.

\FigRatiosRthresAlpha{}

\subsection{Potential degeneracy: line ratio variations with $R_{\rm{thresh}}$ \& $\sigma$}

Figure~\ref{fig:emission-function:rthres-sig0} shows the line ratios as a function of $r_{\rm{thres}}$ and $\sigma$, for a fixed LN mean value of $N_0 = 5\times10^{21}$\,cm$^{-2}$ and a fixed PL slope $\alpha = 2.5$.

The $r_{\rm{thres}}$ parameter is notably degenerate. For $r_{\rm{thres}}$$>10$, there is little to no variation of the line ratios with this parameter. As before, as $r_{\rm{thres}}$ increases, the PL portion of $N$-PDF becomes negligible and its parameters become intractable.
For $r_{\rm{thres}}$$<10$, there is a degeneracy between $\sigma$ and $r_{\rm{thres}}$, as shown by the isocontours curved pattern. This case of degeneracy is illustrated in Figure~\ref{fig:degeneracies:sigma-rthres}. In the limit case where $r_{\rm{thres}}$$<\sim2$, the LN width parameter $\sigma$ is degenerate. In this case the $N$-PDF is almost entirely PL, and the LN part of the $N$-PDF shapes lower densities only where the emission function is close to null. This type of degeneracy is illustrated in Figure~\ref{fig:degeneracies:sigma}.

\FigRatiosRthresSig{}

\subsection{Potential degeneracy: line ratio variations with $\alpha$ \& $\sigma$}

Figure~\ref{fig:emission-function:alpha-sig0} shows the line ratios as a function of $\alpha$ and $\sigma$, for a fixed LN average value of $N_0$ = 5$\times$10$^{21}$\,cm$^{-2}$ and a fixed $r_{\rm{thres}}$ = 1.1.

The $\sigma$ parameter is totally degenerate. This is caused by the rather low average density $N_0$, combined with an early PL onset $r_{\rm{thres}}$ = 1. In this case, the density distribution is dominated by the PL. Variations of $\sigma$ result in variation of the PDF shape for $N_{\rm{H_2}}$ < 1.1$N_0$, however at these column densities the emission of lines is near 0, therefore variations of $\sigma$ do not produce intensity variations.

\FigRatiosAlphaSig{}






\section{Emission function parameters and validity intervals of the global emission model}

\begin{table}[h]
    \centering
    \caption{Parameters of the smoothly broken power law function (equation~\ref{eq:bkn_pl}) fitted on the Orion~B data through $\chi^2$ minimisation in Section~\ref{sec:model:emission-function}.}
    \begin{tabular}{llllll}
    \hline
        Line & \modifr{$f_{l,b}$} & $N_b$ & $\beta_1$ & $\beta_2$ & $\Delta$  \\
        \hline
$^{12}$CO(1-0) & $1.02 \times 10^{-05}$ & $4.83 \times 10^{+20}$ & 46.13 & 0.30 & 0.48 \\ 
$^{13}$CO(1-0) & $4.10 \times 10^{+00}$ & $2.86 \times 10^{+21}$ & 4.79 & 0.55 & 0.31 \\ 
C$^{18}$O(1-0) & $4.14 \times 10^{-24}$ & $2.19 \times 10^{+20}$ & 103.00 & 0.53 & 0.75 \\ 
HCO$^{+}$(1-0) & $1.00 \times 10^{+00}$ & $4.35 \times 10^{+21}$ & 2.73 & 0.57 & 0.64 \\ 
HCN(1-0) & $1.68 \times 10^{+00}$ & $5.99 \times 10^{+21}$ & 2.30 & 0.91 & 0.05 \\ 
HNC(1-0) & $9.25 \times 10^{-24}$ & $2.14 \times 10^{+20}$ & 102.63 & 0.79 & 0.71 \\ 
$^{12}$CS(2-1) & $7.22 \times 10^{-01}$ & $6.25 \times 10^{+21}$ & 2.32 & 1.15 & 0.08 \\ 
SO(3$_2$-2$_1$) & $1.47 \times 10^{-01}$ & $4.21 \times 10^{+21}$ & 4.99 & 0.90 & 0.40 \\ 
N${_{2}}$H$^{+}$(1-0) & $2.02 \times 10^{+00}$ & $2.58 \times 10^{+22}$ & 3.14 & 0.92 & 0.14 \\ 
\hline
    \end{tabular}
    \label{tab:bkn_pl_params}
\end{table}

\begin{table}[h]
\centering
\caption{Limits of the validity interval and prior distribution type of the inverted model parameters.}
\begin{tabular}{lccl}
\hline
\textbf{Parameter} & \textbf{Min.} & \textbf{Max.} & \textbf{prior distribution} \\ \hline
$N_0$             & 10$^{20}$                & 10$^{23}$               & uniform (log)                   \\ 
$\sigma$            & 0.1                & 2               & uniform (log)                   \\ 
$r_{\rm{thres}}$             & 1.1                & 10$^{5}$              & uniform (log)               \\ 
\modifr{$d_\alpha$}             &    \modifr{0.05}           & \modifr{1}               & uniform (log)    \\
\modifr{$\eta$}             &    \modifr{0.01}           & \modifr{1}               & uniform (log)  
\\ \hline
\end{tabular}
\label{tab:val_intervals}
\end{table}

\end{appendix}
\end{document}